\documentclass[aps, twocolumn, preprintnumbers, amsmath, amssymb, amsfonts, superscriptaddress, nofootinbib]{revtex4-2}
\ifdefined\pdfoutput\else\let\pdfoutput\outputmode\fi
\pdfoutput=1

\usepackage{amsmath,amssymb,bm,mathrsfs}
\usepackage{hyperref}
\usepackage{url}
\usepackage{graphicx}
\usepackage{xcolor}
\usepackage{tikz}
\usepackage[compat=1.1.0]{tikz-feynman}
\usepackage[normalem]{ulem}
\usepackage{braket}
\usepackage{appendix}
\usepackage{MnSymbol}
\usepackage{slashed}
\usepackage{comment}

\definecolor{Orange}{cmyk}{0,0.50,1,0}
\definecolor{SkyBlue}{cmyk}{0.80,0,0,0}
\definecolor{BluishGreen}{cmyk}{0.97,0,0.75,0}
\definecolor{Yellow}{cmyk}{0.10,0.05,0.90,0}
\definecolor{Blue}{cmyk}{1,0.50,0,0}
\definecolor{Vermillion}{cmyk}{0,0.80,1,0}
\definecolor{ReddishPurple}{cmyk}{0.10,0.70,0,0}
\definecolor{Black}{cmyk}{0,0,0,1}

\newcommand{\mr}[1]{\mathrm{#1}}
\newcommand{\mc}[1]{\mathcal{#1}}
\newcommand{\tx}{\text}

\renewcommand{\ol}{\overline}
\newcommand{\paren}[1]{\left (#1\right) }
\newcommand{\sqbr}[1]{\left [#1\right] }
\newcommand{\pn}{\paren}
\newcommand{\autospace}{%
  \mathchoice%
    {\!}
    {\!}
    {}
    {}
}
\newcommand{\fn}[1]{\autospace\paren{#1}} 

\newcommand{\SD}{\tx{R}}

\newcommand{\bs}[1]{\boldsymbol{#1}}

\newcommand{\sumSorD}{\sum_{\mc N_{\SD\pm}}}
\newcommand{\sumSandD}{\sum_{\SD = \tx{S}, \tx{D}}}
\newcommand{\sumIandJ}{\sum_{X=I,J}}
\newcommand{\sumIandJprime}{\sum_{X'=I,J}}
\newcommand{\al}[1]{\begin{align}#1\end{align}}
\newcommand{\DTexp}{{\cal I}}

\DeclareMathOperator{\im}{Im}
           
\begin{document}
\hfill TU-1309

\title{Real and Virtual Propagation in Neutrino Oscillations}

\author{Kenji Nishiwaki}
\email{kenji.nishiwaki@snu.edu.in}
\affiliation{Department of Physics, School of Natural Sciences, Shiv Nadar Institution of Eminence (Deemed to be University),
Tehsil Dadri, Greater Noida, Uttar Pradesh, 201314, India}

\author{Kin-ya Oda}
\email{odakin@lab.twcu.ac.jp}
\affiliation{Department of Information and Mathematical Sciences, 
Tokyo Woman's Christian University, 
Tokyo 167-8585, Japan}

\author{Juntaro Wada}
\email{juntaro.wada.e5@tohoku.ac.jp }
\affiliation{Department of Physics, Tohoku University, Sendai, Miyagi 980-8578, Japan}
\affiliation{Technical University of Munich (TUM), School of Natural Sciences, Physics Department, James-Franck-Str. 1, 85748 Garching, Germany,
}

\begin{abstract}\noindent
We revisit flavor oscillations in vacuum in terms of the propagation time of intermediate states. In the limit of a long propagation time  (or distance), degenerate intermediate states exhibit oscillatory behavior, as described by the Jacob--Sachs (or Grimus--Stockinger) theorem within wave-packet quantum field theory. 
By explicitly evaluating the relevant integrals using the saddle-point method, we derive an extended expression for the flavor-changing amplitude that remains valid even for shorter propagation times.
We show that oscillations occur only when the propagation time exceeds a threshold set by the energy uncertainty of the external wave packets and by the decay width of the propagating particle. For shorter propagation, the intermediate particle 
behaves as a purely virtual state, in the sense that it cannot propagate over a macroscopic distance.
Although a direct experimental test of the transition from virtual to real propagation is challenging, since it typically occurs at microscopic scales, our result implies that the Jacob--Sachs theorem holds to higher accuracy than previously expected, even at short propagation times. Our formalism 
applies not only to neutrinos but also to other propagating particles, and future improvements in energy resolution may 
make this threshold observable.
\end{abstract}

\maketitle

\section{Introduction}

Neutrino flavor oscillation~\cite{Pontecorvo:1957cp, Pontecorvo:1957qd, Katayama:1962mx, Maki:1962mu} (see also~\cite{Gribov:1968kq, Eliezer:1975ja, Fritzsch:1975rz, Bilenky:1976cw, Bilenky:1975tb, Bilenky:1976yj}) is one of the most interesting quantum phenomena in particle physics, and at the same time embodies deep conceptual aspects of quantum theory. 
Following 
its experimental discovery~\cite{Super-Kamiokande:1998kpq, Super-Kamiokande:2004orf, SNO:2001kpb, SNO:2002tuh, SNO:2002hgz} (see e.g.,~\cite{Esteban:2024eli, JUNO:2025gmd} for the latest information), 
it has continued to attract considerable attention, both as a sensitive probe of possible new physics beyond the Standard Model (BSM) and as a test of the Standard Model (SM) framework~\cite{ParticleDataGroup:2024cfk}.

From a theoretical perspective, the wave-packet formulation in quantum field theory (WPQFT) provides one of the most fundamental frameworks for describing neutrino oscillations, 
and the basics of the formalism were established in~\cite{Giunti:1993se, Grimus:1996av, Giunti:1997sk, Grimus:1998uh, Giunti:1997wq, Cardall:1999ze, Grimus:1999ra, Beuthe:2001rc}.
(See also~\cite{Nussinov:1976uw,Kayser:1981ye,Kobzarev:1981ra, Giunti:1991ca,Giunti:1991sx,Rich:1993wu,Kiers:1995zj,Grossman:1996eh,Campagne:1997fu,Kiers:1997pe,Burkhardt:1998zj,Ioannisian:1998ch,Giunti:2000kw,Beuthe:2002ej} as early works before the notable review~\cite{Beuthe:2001rc} (also refer to~\cite{Zralek:1998rp,Lipkin:1999nb,Giunti:2007ry,Xing:2011zza});
for further information on subsequent theoretical developments, refer to, e.g.,~\cite{Giunti:2002xg, Lipkin:2003hj, Garbutt:2003ih, Asahara:2004mh, Fujii:2004px, Nishi:2005dc, Bernardini:2006ak, Boyanovsky:2007zz, Visinelli:2008ds, Cohen:2008qb, Bernardini:2010zba, Akhmedov:2009rb, Akhmedov:2010ms, Wu:2010yr, Wu:2010tr, Naumov:2010um, Akhmedov:2010ua, Boyanovsky:2011xq, Akhmedov:2012uu, Naumov:2013bea, Naumov:2013uia, Hansen:2016klk, Karlovets:2016jrd, Akhmedov:2017xxm, Kobach:2017osm, Stirner:2018ojk, Akhmedov:2019iyt, Egorov:2019vqv, Falkowski:2019kfn, Grimus:2019hlq, Karlovets:2020odl, Naumov:2020yyv, Naumov:2022kwz, Cheng:2022lys, Karamitros:2022nnh, Kovalenko:2022goz, Grimus:2023ktd, Mitani:2023hpd, Dadic:2023tuc, Dobrev:2025crc, Delepine:2026owb}.)\footnote{
Another approach to neutrino oscillations based on quantum field theory focuses on the Bogoliubov transformation (see e.g.,~\cite{Blasone:1998hf, Blasone:1999jb, Smaldone:2021mii, Blasone:2025atj}). This paper will not cite references on developments in neutrino oscillations, which, while interesting, are not directly relevant to the main argument. These include formalisms for describing oscillations beyond the vacuum background and information-theoretic analyses.}

A widely used realization of WPQFT is the so-called external wave-packet model/formalism, in which the external particles participating in the production and detection processes are represented by localized wave packets~\cite{Kayser:1981ye, Giunti:1993se, Grimus:1996av, Giunti:1997sk, Grimus:1998uh, Beuthe:2001rc}. These wave packets encode the spacetime
localization of the source and detection regions associated with the oscillating intermediate particle.
This framework allows one to describe physical observables that explicitly depend on the propagation distance and time of the internal mediating particle, which cannot be captured within the conventional plane-wave description in quantum field theory (QFT).\footnote{
In actual physical processes, neutrino oscillations are described as transitions of intermediate-state neutrinos, since 
the intermediate neutrino is not directly measured. Thus, the description of transition neutrinos as off-shell states in QFT is natural. Furthermore, if energy and momentum conservation are perfectly maintained at the interaction points of the overall physical process, the masses of the intermediate states can be determined from the information of the external line, and neutrino oscillations should cease to exist. The description of external states as wave packets inherently resolves this issue.}
Thus, it has been widely applied both to phenomenological issues such as flavor oscillations and decoherence, and to conceptual questions, including the relation between quantum-mechanical and QFT descriptions~\cite{Beuthe:2001rc, Giunti:2002xg,  Akhmedov:2009rb, Akhmedov:2010ms, Akhmedov:2019iyt}.

Within this context, the Jacob--Sachs theorem~\cite{Jacob:1961zz} plays a central role. It shows that, in the limit of 
a sufficiently long propagation time, the intermediate particle effectively behaves as a real particle and exhibits oscillatory behavior~\cite{Beuthe:2001rc}.\footnote{
As an alternative way to extract the asymptotic form, one may use a method based on the Grimus--Stockinger theorem~\cite{Grimus:1996av}. This approach assumes a sufficiently long propagation distance and suggests the asymptotic behavior of a real particle. 
} 
This theorem is practically useful for understanding flavor oscillations and is consistent with physical intuition; however, two conceptual issues remain in this asymptotic treatment:
\begin{itemize}
    \item First, between the source and detection regions, the intermediate state must probe the pole of the propagator to behave as a real particle. Since the Jacob--Sachs theorem~\cite{Jacob:1961zz} describes only asymptotic states, it does not clarify how a particle produced as a virtual state kinematically reaches the pole and becomes a real particle.
    \item Second, for shorter propagation times, it is not clear how the asymptotic states derived from the Jacob--Sachs theorem~\cite{Jacob:1961zz} are modified. Although the magnitude of such corrections is known to be small in typical experimental setups for SM particles that exhibit oscillatory behavior (e.g., the propagation of neutrinos and neutral $K$ and $B$ mesons)~\cite{Jacob:1961zz, Beuthe:2001rc}, their explicit form has not been derived. However, these corrections may be particularly relevant for flavor transitions of BSM particles.
\end{itemize}

In this work, we address these issues within the Gaussian WPQFT framework developed in~\cite{Ishikawa:2005zc, Ishikawa:2018koj, Ishikawa:2020hph, Ishikawa:2021bzf, Ishikawa:2023bnx, Oda:2024keo}, with manifestly parametrized most probable configurations of the momenta and positions of external states around the initial or final time of quantum transitions.
By directly evaluating the relevant integrals using the saddle-point method, we derive an expression for the flavor-transition amplitude that is valid beyond the asymptotic regime. We clarify the conditions under which the intermediate particle behaves as a real propagating state and oscillates, then identify a threshold propagation time that separates real and virtual regimes. 
Although our analysis is motivated by neutrino oscillations, the formalism is applicable to more general oscillating particles.

This paper is organized as follows.
In Sec.~\ref{sec: Neutrino Oscillation in WPQFT}, we review the conventional treatment of neutrino flavor oscillations in WPQFT, including the Jacob--Sachs theorem. In Sec.~\ref{sec: Real and Virtual Propagation of intermediate particles}, we go beyond the asymptotic approximation and evaluate the propagation amplitude using the saddle-point method. Then, we clarify the existence of the threshold 
propagation time that separates virtual and real propagation.
In Sec.~\ref{sec: Flavor Changing Probability}, we show the expression 
for the flavor-changing probability and discuss whether we can probe the threshold time in the current experimental setup.
Finally, Sec.~\ref{sec: Discussions and Summary} presents discussions and a summary of our findings.
In App.~\ref{app:brief-summary}, we provide a summary of the key points regarding the Gaussian WPQFT that we employ, along with 
the notation used.
In App.~\ref{app: Single-field Gaussian state and Contraction}, we summarize the rules for the contraction between fields and external-line states represented by wave packets in the Gaussian WPQFT formalism.
In App.~\ref{app: Towards an Effective Propagator}, we present an explicit derivation of the expression for the effective propagator.
In App.~\ref{app: Evaluation of the internal-momentum integration}, we provide details regarding the calculation of the four-dimensional momentum integral for intermediate states.
In App.~\ref{app: Evaluation of the propagating time integration}, we explain how to derive the transition probability of neutrino flavors in an 
approximate way under phenomenological assumptions.

\section{Neutrino Oscillation in WPQFT}
\label{sec: Neutrino Oscillation in WPQFT}
In this section, we review the conventional treatment of flavor oscillations within WPQFT. We provide the amplitude in WPQFT computation, its evaluation using the Jacob--Sachs theorem, and the derivation of the flavor-changing probability.

\subsection{External Wave Packet Model}
\label{sec: External Wave Packet Model}

In QFT treatments of neutrino oscillations, the propagation time and distance are introduced by requiring the particles at the production and detection vertices to be sufficiently localized and separable~\cite{Kayser:1981ye, Giunti:1993se, Grimus:1996av, Giunti:1997sk, Grimus:1998uh, Beuthe:2001rc}. We refer to these as the source region S and the detection region D, respectively. Localization is implemented by assigning Gaussian wave packets to all external lines and performing the calculation accordingly~\cite{Giunti:1993se, Grimus:1996av, Giunti:1997wq, Giunti:1997sk, Grimus:1999ra}.
A framework based on these assumptions is known as the external wave packet model~\cite{Beuthe:2001rc}.

We consider a production process in which a single neutrino is generated in S,
\begin{align}
\label{eq: interaction at source region}
i_{\tx S+} + j_{\tx S+} + \cdots
    \to \nu_{\alpha} + i'_{\tx S-} + j'_{\tx S-} + \cdots,
\end{align}
and a detection process in which the produced neutrino is observed in D,
\begin{align}
\label{eq: interaction at detection region}
\nu_{\beta} + a_{\tx D+} + b_{\tx D+} + \cdots
    \to a'_{\tx D-} + b'_{\tx D-} + \cdots,
\end{align}
where $\alpha$ and $\beta$ represent the corresponding neutrino flavors, which in principle can differ.
The subscript $\tx{S}$ or $\tx{D}$ on each particle indicates its region, hereafter denoted generically by the letter~$\SD$,
\al{
\SD	&=	\tx{S},\tx{D}.
	\label{R is S or D}
}%
The signs $\pm$ distinguish incoming ($+$) from outgoing ($-$) particles, i.e., the initial and final states of the entire quantum transition process.
Here, we describe the neutrino as an internal line, since it is not directly observed. Consequently, in WPQFT, the neutrino production and detection are treated not as two separate processes but as a single unified one.

\begin{figure}[h]
\centering
\begin{tikzpicture}
\begin{feynman}
  \vertex (vS) at (0,0);
  \vertex [left=1cm of vS] (nS) {\(\tx{n}_{\tx S+}\)};
  \vertex [right=3cm of vS] (pS) {\(\tx{p}_{\tx{S}-}\)};
  \vertex [above right=2cm of vS] (eS) {\(\tx{e}^{-}_{\tx{S}-}\)};
  \vertex (vD) at (3.5,-3.2);
  \vertex [left=2cm of vD] (pD) {\(\tx{p}_{\tx{D}+}\)};
  \vertex [below right=2cm of vD] (eD) {\(\tx{e}^{+}_{\tx{D}-}\)};
  \vertex [right=2cm of vD] (nD) {\(\tx{n}_{\tx{D}-}\)};

  \diagram* {
    (nS) -- [fermion] (vS),
    (vS) -- [fermion] (pS),
    (vS) -- [fermion] (eS),
    (vS) -- [anti fermion, edge label=\(\ol{\nu}_e\)] (vD),
    (pD) -- [fermion] (vD),
    (eD) -- [fermion] (vD),
    (vD) -- [fermion] (nD),
  };
\end{feynman}
  \fill (vS) circle (2.4pt);
  \fill (vD) circle (2.4pt);
  \node at (-0.45,0.4) {\(\tx{S}\)};
  \node at (3.95,-2.75) {\(\tx{D}\)};
  \draw[->, line width=0.9pt, black!55] (1,-6.0) -- (3,-6.0);
  \node[black!55, anchor=south] at (2,-5.95) {time};
\end{tikzpicture}
\caption{Production and detection processes of an electron antineutrino. The dots labeled $\tx{S}$ and $\tx{D}$ mark the source (production) and detection interaction points, respectively.}
\label{fig: source and detection diagram beta decay}
\end{figure}
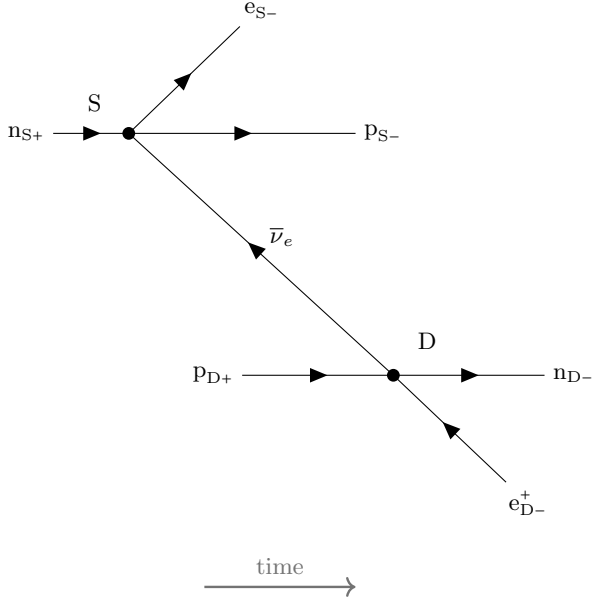

As a simple illustrative example of this sequence of processes, we consider the production of an electron antineutrino via $\beta$ decay,
\begin{align}
\label{eq: source process}
\tx{n}_{\tx S+} \to \ol{\nu}_e + \tx{p}_{\tx S-} + \tx{e}^{-}_{\tx S-},
\end{align}
and the detection process via its inverse decay:
\begin{align}
\label{eq: detection process}
\ol{\nu}_e + \tx{p}_{\tx D+} \to \tx{e}^{+}_{\tx D-} + \tx{n}_{\tx D-}.
\end{align}
Here $\tx{n}_{\tx S +}$, $\tx{p}_{\tx{S}-}$, and $\tx{e}^{-}_{\tx{S}-}$ represent the incoming neutron, outgoing proton, and outgoing electron 
involved in the production process, while $\tx{p}_{\tx{D}+}$, $\tx{e}^{+}_{\tx{D}-}$, and $\tx{n}_{\tx{D}-}$ represent the incoming proton, outgoing positron, and outgoing neutron 
involved in the detection process, respectively. We show a Feynman diagram of this process in Fig.~\ref{fig: source and detection diagram beta decay}.

To distinguish particles participating in the sequence of processes, we introduce the label representation $\mc{N}_{\SD\pm}$.
For instance, for the set of processes~\eqref{eq: source process} and \eqref{eq: detection process}, we have
\begin{align}
\mc N_{\tx{S}\pm} &= \tx{n}_{\tx{S} +}, \tx{p}_{\tx{S}-}, \tx{e}^{-}_{\tx{S}-},&
\mc N_{\tx{D}\pm} &= \tx{p}_{\tx{D}+}, \tx{e}^{+}_{\tx{D}-}, \tx{n}_{\tx{D}-},
\end{align}
where the $\pm$ symbol attached to ${\cal N}_\tx{S}$ or ${\cal N}_\tx{D}$ serves as a reminder that we retain both the `$+$' and `$-$' information inside the representation labels.

Throughout the paper, we adopt the standard notation for any spacetime vector~$V$, with spatial part~$\bs V$:
\al{
V	&=	\pn{V^\mu}_{\mu=0,\dots,3}
	=	\pn{V^0,\bs V},\\
\bs V
	&=	\pn{V^i}_{i=1,2,3}
	=	\pn{V^1,V^2,V^3}.
}
Here and hereafter, Greek (Latin) letters $\mu,\nu,\dots$ ($i,j,\dots$) denote spacetime (spatial) indices running over $0,\dots,3$ ($1,2,3$), and we adopt the Einstein summation convention for them.
We also adopt the standard notation for the Dirac delta function (distribution) $\delta^4\fn{V}:=\prod_{\mu=0}^3\delta\fn{V^\mu}$.

In this model, the external lines $\Ket{\mc N_{\SD+}}$ and $\Bra{\mc N_{\SD-}}$ are created from the vacuum by the wave-packet creation operators of $\widehat{\psi}$, and their contraction with $\widehat{\psi}$ is given by
\begin{align}
\widehat{\psi} \fn{x}\Ket{\mc N_{\SD+}}
	\sim 
    \int d^3 \bs{p}\,\exp\!\sqbr{G_{\mc{N}_{\SD+}}\fn{x, \bs{p}}}, \notag \\
\Bra{\mc N_{\SD-}} \overline{\widehat{\psi}}
    \sim 
    \int d^3 \bs{p}\,\exp\!\sqbr{G_{\mc{N}_{\SD -}}\fn{x, \bs{p}}},
    \label{eq:psihat-Nstate}
\end{align}
where $G_{\mc N_{\SD \pm}}\fn{x, \bs p}$ is the Gaussian factor
\begin{align}
\label{eq: Gaussian exponent}
G_{\mc N_{\SD \pm}}\fn{x, \bs p}
	&:=	-{\sigma_{\mc N_{\SD \pm}} \over 2}\pn{\bs{p}-\bs{P}_{\mc N_{\SD \pm}}}^2 \nonumber \\
    &\quad
		\mp iE_{\psi}\fn{\bs p}\pn{x^0-X_{\mc N_{\SD \pm}}^0}
        \pm i\bs p\cdot\pn{\bs x-\bs X_{\mc N_{\SD \pm}}},
\end{align}
and we have omitted the spinor wave function and the normalization factor.

Note that
\begin{align}
\bs X_{\mc N_{\SD \pm}}, 
\bs P_{\mc N_{\SD \pm}}, \sigma_{\mc N_{\SD \pm}}
    \label{eq:external-variables}
\end{align}
represent the central position, the central momentum, and the width-squared of the wave packet for the corresponding external particle at time $X^0_{\mc N_{\SD \pm}}$. Here, $\sigma_{\mc N_{\SD \pm}}$ has mass dimension $[\sigma_{\mc N_{\SD \pm}}] = -2$, corresponding to the inverse squared momentum uncertainty.
In Appendix~\ref{app: Single-field Gaussian state and Contraction}, we collect the exact expressions for the incoming and outgoing particle states, including the spinor factors, as well as their forms after evaluating the external momenta at the saddle point.

Since the sign of the momentum phase differs for incoming and outgoing particles, the corresponding sign labels have been assigned.
The distinction between particles associated with the source region and those associated with the detection region will also become relevant, as discussed later. By contrast, the difference between particles and antiparticles is unimportant for the behavior of the exponential factor, and we therefore omit labels distinguishing them for notational simplicity. The on-shell energy of a particle is denoted by $E_{\psi}$,
\begin{align}
E_{\psi}\fn{\bs{p}} := \sqrt{m_{\psi}^2 + \bs{p}^2},
\label{eq:def_energy}
\end{align}
where $m_{\psi}$ is the mass of the particle associated with the field $\psi$ that creates the state $\Ket{\mc N_{\SD \pm}}$. The corresponding velocity $v^i_{\psi}\fn{\bs{p}}$ is given by 
\begin{align}
v^i_{\psi}\fn{\bs{p}}
	:={p^i \over E_{\psi}\fn{\bs{p}}}.
    \label{eq:def_velocity}
\end{align}

\subsection{Amplitude in WPQFT}
\label{subsec: Amplitude in WPQFT}
In the external wave-packet model with Gaussian profiles, the evaluation of the amplitude for flavor-changing (or flavor-conserving) processes, $\mc A_{I,\alpha\beta} := \mc A_I\fn{\nu_\alpha \to \nu_\beta}$, involves integrals over the external momenta, the interaction spacetime points, and the propagating momentum.
Here and hereafter, the indices $I,J,\dots$ label the intermediate neutrino mass eigenstates.

Since Gaussian wave packets break Lorentz invariance, closed-form analytic expressions are generally inaccessible.\footnote{
This issue may be partially resolved by employing Lorentz-covariant wave packets. For related discussions, we refer to Refs.~\cite{Naumov:2009zza, Naumov:2010um, Naumov:2020yyv, Oda:2021tiv, Oda:2023qek}.
}
We therefore evaluate the relevant integrals using the saddle-point method, valid for sharply peaked (large-$\sigma$) wave packets, within the Gaussian WPQFT framework~\cite{Ishikawa:2018koj, Ishikawa:2020hph, Ishikawa:2021bzf, Ishikawa:2023bnx}.

Using the saddle-point method to carry out all integrations except the one over the propagating momentum, we are left with an integral of the following form~\cite{Beuthe:2001rc}:
\begin{align}
\mc A_{I, \alpha\beta}
	&\simeq
			 \Bigg(
			\int\frac{d^4 p_{\nu_I}}{\pn{2\pi}^4}
					  G\fn{p^0_{\nu_I}, \bs{p}_{\nu_I}}
                      \Psi\fn{p^0_{\nu_I}, \bs{p}_{\nu_I}}  \nonumber\\
    &\phantom{\simeq\Bigg(\int\frac{d^4 p_{\nu_I}}{\pn{2\pi}^4}}
			e^{-ip_{\nu_I}^0\Delta T\,+\,i\bs p_{\nu_I}\cdot\bs L\,+\,i\varphi}
					\Bigg) \nonumber \\
    &\quad \times
			M_{I,\alpha\beta}
            \fn{m_{\nu_I}, \, p^0_{\nu_I},\bs{p}_{\nu_I}}.
\label{eq: effective propagator}
\end{align}
Here, the quantities appearing above are defined as follows:
\begin{itemize}
\item $G\fn{p^0_{\nu_I}, \bs{p}_{\nu_I}}$ represents the scalar part of the propagator,
\begin{align}
\label{eq: neutrino propagator}
G\fn{p^0_{\nu_I}, \bs{p}_{\nu_I}}
	:=
        \frac{-i }{p_{\nu_I}^2+m_{\nu_I}^2-i m_{\nu_I} \Gamma_{\nu_I}},
\end{align}
where $p_{\nu_I}$ is the off-shell propagating momentum and $\Gamma_{\nu_I}$ is the decay width of $\nu_I$. 
Until we derive the probability formula, we use $\nu_I$ as a generic label for either an active or a sterile neutrino.
\item The function
\begin{align}
\Psi\fn{p^0_{\nu_I}, \bs{p}_{\nu_I}}
\end{align}
is referred to as the overlap function and encodes the localization properties of the wave packets. Its explicit form for Gaussian wave packets is derived in Appendix~\ref{app: Towards an Effective Propagator}. In the present section,
we require only the following property: $\Psi$ is finite when $p^0_{\nu_I} > 0$ and when $-p_{\nu_I}^2 = (p_{\nu_I}^{0})^2 - \bs{p}_{\nu_I}^{\,2}$ lies within a range of order the inverse width-squared of the wave packet, while $\Psi$ is negligibly small outside this region.
\item 
The quantities $\Delta T$ and $\bs L$ form a Lorentz vector $\pn{\Delta T,\bs L}$ representing the spacetime distance of the propagating intermediate particle:
\begin{align}
&&
\Delta T
	&:=	T_{\tx{D}} - T_{\tx{S}},&
L^i
	&:=	X^i_{\tx{D}}-X^i_{\tx{S}},
\label{eq:Delta-variables}
\end{align}
where $T_{\SD}$ is the real part of the saddle point of the interaction time, and $\bs{X}_{\SD}$ is that of the interaction spatial coordinate, evaluated at the corresponding time; recall Eq.~\eqref{R is S or D} for the label $\SD$.
Their explicit expressions are shown at the end of this subsection, that is, in Eqs.~\eqref{eq: TSD} and \eqref{eq: XSD} as functions of the external wave-packet variables~\eqref{eq:external-variables}.
In Appendix~\ref{app: Towards an Effective Propagator}, we provide the explicit derivation of $T_{\SD}$ and $\bs{X}_{\SD}$.
We will also write $L:=|\bs{L}|$.
\item Flavor-dependent matrix
\begin{align}
M_{I, \alpha\beta}\fn{m_{\nu_I},\, p^0_{\nu_I},\, \bs{p}_{\nu_I}}
\end{align}
factorizes out the vertex contributions, including the flavor-mixing matrix, namely the Pontecorvo--\allowbreak(Katayama--\allowbreak Matumoto--\allowbreak Tanaka--\allowbreak Yamada--)\allowbreak Maki--\linebreak Nakagawa--\allowbreak Sakata (P[KMTY]MNS) matrix~\cite{Pontecorvo:1957cp, Pontecorvo:1957qd, Katayama:1962mx, Maki:1962mu} for SM neutrinos, together with the spinor functions. 
\item 
The phase $\varphi$ is momentum-independent and therefore drops out upon squaring the amplitude, so it does not affect the transition probability.
\end{itemize}

We comment on the decay width $\Gamma_{\nu_I}$.
For an SM neutrino, the dominant decay channel is 
$\nu_I \to \nu_J + \gamma$ 
through the magnetic moment induced by one-loop interactions~\cite{Shrock:1974nd, Petcov:1976ff, Zatsepin:1978iy}, 
for which~\cite{Giunti:2024gec}
\begin{align}
\Gamma_{\nu_I} \sim 10^{-60}~\mr{eV} .
\end{align}
This value is extremely small.
The decay width of a sterile neutrino $N_I$ into the SM leptons can be estimated, for $m_{N_I} \gg m_{e}$, as
\begin{align}
\Gamma_{N_I} 
	\sim {1 \over 192\pi^3} \left|V_{l_{\alpha} N_I}\right|^2 G_\tx{F}^2 m_{N_I}^5 .
\end{align}
Here, $G_\tx{F}$ is the Fermi constant, and $V_{l_{\alpha} N_I}$ denotes the mixing matrix element between the SM neutrino and the sterile neutrino. Although the precise value depends on the sterile neutrino mass range, a typical constraint is $|V_{e N_1}|^2 \lesssim 10^{-4}$~\cite{Bolton:2022pyf}. Using this value, one finds that if the sterile neutrino couples only to the SM,
\begin{align}
\Gamma_{N_I} 
	\lesssim 10^{-18} \left( \frac{m_{N_I}}{1 ~\mr{GeV}} \right)^5~\mr{GeV} .
    \label{eq:Gamma-sterile-bound}
\end{align}

For later convenience, we introduce $\sigma_{\SD}$, whose  inverse square root represents the averaged  momentum uncertainty of the external wave packets 
\begin{align}
\sigma_{\SD} := \left( \sumSorD \sigma^{-1}_{\mc N_{\SD\pm}}\right)^{-1};
\label{eq:averaged-signma}
\end{align}
recall Eq.~\eqref{R is S or D} for the label \SD.
A weighted average for the quantity $C$ with order $k$ is defined as\footnote{
The overline here denotes the weighted average and should not be confused with the Dirac adjoint in Eq.~\eqref{eq:psihat-Nstate}.
}
\begin{align}
\ol{C}_{\SD k}
    &:= \sigma_\SD^k \sumSorD \frac{(\pm)^{C} \, C_{\mc{N}_{\SD\pm}}}{\left(\sigma_{\mc{N}_{\SD\pm}}\right)^k
    		}
    \label{overline notation}
\end{align}
for $\SD=\tx{S},\tx{D}$ as in Eq.~\eqref{R is S or D}, where $(\pm)^{C}$ denotes a sign factor that depends on the type of parameter $C_{\mc{N}_{\SD}\pm}$:
\begin{align}
(\pm)^{C}
    :=
\begin{cases}
\pm & (\text{if } C_{\mc{N}_{\SD}\pm} \text{ contains an odd number of } P_{\mc{N}_{\SD\pm}}^\mu),\\
1 & (\text{otherwise}).
\end{cases}
    \label{eq:pmC-operation}
\end{align}

For example, the weighted average of the central momentum, the time-dependent central position, and the velocity is given by
\begin{align}
\ol{P^i}_{\SD 0} 
    &= \sumSorD \pm P^i_{\mc{N}_{\SD\pm}}, \\
\ol{X_{\medstar}^i \fn{x^0}}_{\SD1}
    &= \sigma_{\SD} \sumSorD \sigma^{-1}_{\mc N_{\SD\pm}}\,
        X_{\medstar \mc N_{\SD\pm}}^i\fn{x^0},\\
\ol{v^i}_{\SD 1}
    &= \sigma_\SD
        \sumSorD
        \sigma^{-1}_{\mc N_{\SD\pm}}\,
        v_{\psi}^{i}\fn{\bs{P}_{\mc{N}_{\SD\pm}}}, 
\end{align}
where the $\pm$ sign in $\ol{P^i}_{\SD 0}$ follows from Eq.~\eqref{eq:pmC-operation}, since $P^i$ is odd in momentum. No such sign appears in $\ol{v^i}_{\SD 1}$, since $v^i = P^i/P^0$ contains an even number of momentum factors, counting the denominator $P^0$; see Eq.~\eqref{eq:def_velocity}.
Here, we define the variable,
\begin{align}
X_{\medstar\mc N_{\SD\pm}}^{i}\fn{x^0}
	&:=	X_{\mc N_{\SD\pm}}^{i}+v_{\psi}^{i}\fn{\bs{P}_{\mc N_{\SD\pm}}} \pn{x^0-X_{\mc N_{\SD\pm}}^0},
    \label{eq:def_Xstar}
\end{align}
which represents the classical trajectory of the center of the corresponding wave packet.

The real part of the saddle point of the interaction time introduced above, $T_{\SD}$, as well as the real part of the saddle point of the interaction coordinate, $\bs{X}_{\SD}$, can be expressed in terms of these weighted averages as
\begin{align}
\label{eq: TSD}
T_{\SD}
    &:= \sigma_{t, \SD}
    \frac{
        \ol{v^i}_{\SD1}\ol{X^i_{\medstar}\fn{0}}_{\SD1}
        - \ol{v^i X^i_{\medstar}\fn{0}}_{\SD1}
    }{\sigma_\SD}, \\
\label{eq: XSD}
\bs{X}_{\SD}
    &:= \ol{\bs{X}_{\medstar}\fn{T_{\SD}}}_{\SD1},
\end{align}
where
\begin{align}
\sigma_{t,\SD}
    := {\sigma_\SD \over  \ol{v^2}_{\SD1}-\ol{v^i}_{\SD1}\ol{v^i}_{\SD1}},
    \label{eq:def_sigma_tSD}
\end{align}
whose \emph{inverse square root} corresponds to the \emph{energy uncertainty} in the source and detection regions.

It is also convenient to introduce the sums of these widths in the source and detection regions:
\begin{align}
\sigma_{\tx{S}+\tx{D}}
	&:= 
	\sigma_{\tx{S}} + \sigma_{\tx{D}},
    \label{eq:sigma_SplusD} \\
\sigma_{t, \tx{S}+ \tx{D}}  
	&:= 
		 \sigma_{t, \tx{S}} +  \sigma_{t, \tx{D}}.
     \label{eq:sigma_tSplusD}
\end{align}

\subsection{Jacob--Sachs theorem}
To evaluate the propagating momentum integral of the amplitude reduced to
Eq.~\eqref{eq: effective propagator}, the Jacob--Sachs theorem is useful~\cite{Jacob:1961zz}.\footnote{
As an alternative way to extract the asymptotic form of the propagating-momentum integral in Eq.~\eqref{eq: effective propagator}, one may use a method based on the Grimus--Stockinger theorem~\cite{Grimus:1996av}. 
This approach performs the spatial-momentum integration of the propagating particle first. As in Eq.~\eqref{eq: Jacob-Sachs theorem}, the pole of the propagator provides the dominant contribution. However, extending this method to unstable oscillating particles is technically challenging~\cite{Beuthe:2001rc}.
}
This theorem states that, when the propagation time $\Delta T$ can be regarded as
sufficiently large, the integral over the zeroth component of the propagating momentum,
\begin{align}
I(\Delta T):=
			\int d p^0_{\nu_I}
					  G\fn{p^0_{\nu_I}, \bs{p}_{\nu_I}}
                      \Psi\fn{p^0_{\nu_I}, \bs{p}_{\nu_I}}
			~e^{-i p_{\nu_I}^0 \Delta T} 
\end{align}
admits the following asymptotic behavior~\cite{Jacob:1961zz}:
\begin{align}
\label{eq: Jacob-Sachs theorem}
I(\Delta T) 
	\underset{\Delta T\to\infty}{\longrightarrow} 
	{\pi Z \over \sqrt{z_{\nu_I} + \bs{p}_{\nu_I}^2}} 
    \Psi\fn{\sqrt{z_{\nu_I} + \bs{p}_{\nu_I}^2}, \bs{p}_{\nu_I}}
    e^{-i \sqrt{z_{\nu_I} + \bs{p}_{\nu_I}^2} \Delta T},
\end{align}
where
\begin{align}
z_{\nu_I} := m_{\nu_I}^2 - i m_{\nu_I} \Gamma_{\nu_I},
    \label{eq:z_nu_I-definition}
\end{align}
is the pole of the propagator $ G\fn{p^0_{\nu_I}, \bs{p}_{\nu_I}}$ and $Z$ is the residue with respect to $p_{\nu_I}^2$.

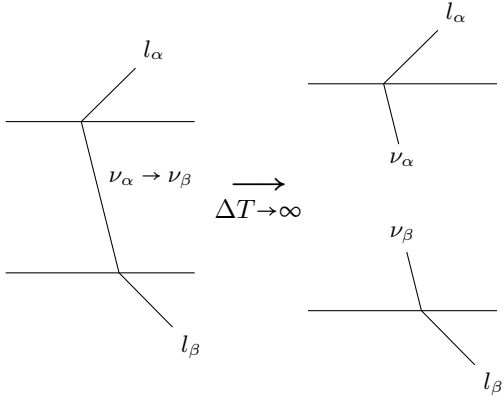
\begin{figure}[h]
\centering
\begin{tikzpicture}
\node (left) at (0,0) {
\begin{tikzpicture}
\begin{feynman}
  \vertex (vS) at (0,0);
  \vertex [left=1cm of vS] (nS) ;
  \vertex [right=1.5cm of vS] (pS) ;
  \vertex [above right=1cm of vS] (eS) 
  {\(l_{\alpha}\)};
  \vertex (vD) at (0.5,-2);
  \vertex [left=1.5cm of vD] (pD) ;
  \vertex [below right=1cm of vD] (eD)
   {\(l_{\beta}\)};
  \vertex [right=1cm of vD] (nD) ;

  \diagram* {
    (nS) -- (vS),
    (vS) -- (pS),
    (vS) -- (eS),
    (vS) -- [edge label=\(\nu_\alpha \to \nu_\beta\)] (vD),
    (pD)--  (vD),
    (eD)--(vD),
    (vD)--(nD),
  };
\end{feynman}
\end{tikzpicture}
};

\node at (2,0) {\Large $\underset{\Delta T \to \infty}{\longrightarrow} $};

\node (right) at (4,0) {
\begin{tikzpicture}
\begin{feynman}
  \vertex (vS1) at (0,0);
  \vertex [left=1cm of vS] (nS) ;
  \vertex [right=1.5cm of vS] (pS) ;
  \vertex [above right=1cm of vS] (eS) 
  {\(l_{\alpha}\)};
  \vertex (vS2) at (0.25,-1) {\(\nu_\alpha\)};
  \vertex (vD1) at (0.25,-2) {\(\nu_\beta\)};
  \vertex (vD2) at (0.5,-3);
  \vertex [left=1.5cm of vD2] (pD) ;
  \vertex [below right=1cm of vD2] (eD) 
  {\(l_{\beta}\)};
  \vertex [right=1cm of vD2] (nD) ;

  \diagram* {
    (nS) -- (vS1),
    (vS) -- (pS),
    (vS) -- (eS),
    (vS1) -- (vS2),
    (vD1) -- (vD2),
    (pD)--  (vD2),
    (eD)--(vD2),
    (vD2)--(nD),
  };
\end{feynman}
\end{tikzpicture}
};
\end{tikzpicture}
\caption{Diagrammatic illustration for Jacob--Sachs theorem.}
\label{fig: Jacob-Sachs theorem}
\end{figure}

This theorem is not only practically useful but also physically instructive. It implies that, when the propagation time is taken to be sufficiently large, only the on-shell (and forward-in-time propagating) contributions of the propagator survive asymptotically. This is reflected in the phase factor in Eq.~\eqref{eq: Jacob-Sachs theorem}, $e^{- i \sqrt{z_{\nu_I} + \bs{p}_{\nu_I}^2}\,\Delta T}$, which asymptotically reduces to the on-shell phase. This implies that the propagator asymptotically approaches the behavior of an external line. For example, the diagram in Fig.~\ref{fig: source and detection diagram beta decay} is decomposed, in the limit of a sufficiently long propagation time $\Delta T \to \infty$, into the two processes~\eqref{eq: source process} and~\eqref{eq: detection process}. A schematic illustration is shown in Fig.~\ref{fig: Jacob-Sachs theorem}.

The corrections to the asymptotic formula~\eqref{eq: Jacob-Sachs theorem} are safely negligible for typical experimental setups involving commonly studied oscillating particles such as neutrinos and neutral $K$ and $B$ mesons~\cite{Jacob:1961zz, Beuthe:2001rc}. This is because experiments usually probe the behavior of particles that have propagated over long times. However, depending on the hadronic species, these corrections can become non-negligible; for example, Ref.~\cite{Beuthe:2001rc} pointed out this possibility for the $\Delta(1232)$ resonance. More generally, for BSM particles, such corrections may be relevant in certain regions of parameter space.

\subsection{Standard Formula for Flavor Changing Probability via Oscillation}
\label{subsec: flavor changing probability}
After applying the Jacob--Sachs theorem, we evaluate the remaining propagating momentum integrals. This is often done using the saddle-point method and similar techniques, but since the integrand no longer contains any poles, the computation can be carried out straightforwardly~\cite{Beuthe:2001rc}. The flavor transition probability is then obtained by integrating the squared amplitude over the propagation time:
\begin{equation}
\label{eq: probability given by time integration}
P(\alpha \to \beta)
	\propto \sum_{IJ} \int d \Delta T \,
	\mc A_{I,\alpha\beta}\,
	\mc A^{*}_{J,\alpha\beta} .
\end{equation}
The necessity of performing the $\Delta T$ integration is explained by the fact that the propagation time is not measured in experiments~\cite{Kiers:1995zj, Giunti:1997sk, Giunti:1997wq, Beuthe:2001rc, Akhmedov:2009rb}.

When the propagating particle is relativistic, the mass dependence contained in 
$M_{I,\alpha\beta}\!\left(m_{\nu_I},\, p^0_{\nu_I},\, \boldsymbol{p}_{\nu_I}\right)$ 
can be neglected. In this regime, the oscillation probability takes the following standard form (in our notation)~\cite{Kayser:1981ye, Giunti:1991ca, Giunti:1997wq, Giunti:1997sk}:
\begin{align}
\label{eq: Standard Formula}
\left.P(\alpha \to \beta)\right|_{L \to \infty}
    &
    \propto
       \sum_{IJ}
       { M_{I,\alpha\beta} M^*_{J,\alpha\beta}
       \over \sum_{KL} M_{K,\alpha\alpha} M^*_{L,\alpha\alpha} }
       \nonumber\\
	&\qquad\times
		\exp\fn{
			- {L \over L^{\mr{dec}}_{IJ}}
			-2\pi i\, \frac{L}{L^{\mathrm{osc}}_{IJ}}
			-\left(\frac{L}{L^{\mathrm{coh}}_{IJ}}\right)^{2}
			},
\end{align}
where 
\begin{align}
L^{\mathrm{osc}}_{IJ}
    &:= {4 \pi |\bs{P}_{\nu}| \over \Delta m_{IJ}^2}, 
      \label{eq:def_Losc} \\
L^{\mathrm{coh}}_{IJ}
    &:= 4{\sqrt{\sigma_{\tx{S}+\tx{D}}} \over \Delta m_{IJ}^2}
        \left|\bs{P}_{\nu}\right|^2
        =
         \frac{\sqrt{\sigma_{\tx{S}+\tx{D}}}\left|\bs{P}_{\nu}\right|}{\pi}
        L^{\mathrm{osc}}_{IJ},
       \label{eq:def_Lcoh} \\
L^{\mr{dec}}_{IJ}
    &:= {2\left|\bs{P}_{\nu}\right| \over m_{\nu_I}\Gamma_{\nu_I} + m_{\nu_J}\Gamma_{\nu_J}},
    \label{eq:def_Ldec}\\
\Delta m_{IJ}^2
    &:= m_{\nu_I}^2 - m_{\nu_J}^2,
\end{align}
and
\begin{align}
M_{I,\alpha\beta} := M_{I, \alpha\beta}\!\left(0,\, E_{\nu_I}\fn{\bs{P}_{\nu}},\, \bs{P}_{\nu}\right)
\end{align}
with $\bs{P}_{\nu}$ being the real part of the saddle point for the propagating momentum, which will be given in the next section.
In Eq.~\eqref{eq: Standard Formula}, we have neglected the effect of the dispersion of the propagating wave packet with the propagation time~\cite{Beuthe:2001rc}.\footnote{
We can find the following correspondence with the momentum uncertainty of the propagating neutrino in Eq.~(129) of~\cite{Beuthe:2001rc} as $\paren{2 \sigma_{\tx{S}+\tx{D}}}^{-1/2}$.
}
We note that in WPQFT calculations, exponential terms independent of $L$ also generally appear~\cite{Beuthe:2001rc}, which depend on the uncertainties in momentum and the mass eigenvalues. Here, however, we have displayed only the $L$-dependent terms that directly affect the observed values.

As is well known, this oscillation formula~\eqref{eq: Standard Formula} describes oscillatory flavor transitions with a characteristic distance scale given by the oscillation length $L^{\mathrm{osc}}_{IJ}$.
When the propagation distance exceeds the coherence length $L^{\mathrm{coh}}_{IJ}$, the wave packets separate from each other, and the flavor transitions terminate~\cite{Nussinov:1976uw, Kayser:1981ye, Giunti:1991ca}. In addition, if the intermediate particle has a finite lifetime, it cannot propagate over an arbitrarily long distance. Once the propagation distance exceeds $L^{\mr{dec}}_{IJ}$, the flavor transition probability converges to zero~\cite{Beuthe:2001rc}.

\section{
Saddle-point analysis and Real/Virtual propagation
}
\label{sec: Real and Virtual Propagation of intermediate particles}
The derivation of the standard oscillation formula in WPQFT~\eqref{eq: Standard Formula} based on the Jacob--Sachs theorem provides a quantum-field-theoretical justification of the quantum-mechanical treatment for the neutrino oscillation and is consistent with the physical intuition that neutrinos propagating over macroscopic distances behave as real particles. Nevertheless, several conceptual issues remain in this framework:
\begin{itemize}
    \item In Fig.~\ref{fig: source and detection diagram beta decay}, the neutrino propagates in a $t$-channel--like manner. How does a particle produced as a virtual neutrino in the $t$-channel-like diagram kinematically reaches the pole and becomes a real particle in the asymptotic formula~\eqref{eq: Jacob-Sachs theorem}?\footnote{
    As a clearer illustration of the correspondence, consider the neutrino M\"ossbauer process~\cite{Visscher:1959kka, Kells:1982rm, Kells:1983iac},
    \begin{align*}
    {}^3\tx{H}_{\tx{S}+}
    	&\to \bar{\nu}_{e} +{}^3\tx{He}_{\tx{S}-} + e_{\tx{S}-}(\tx{bound}) , \\
    \bar{\nu}_{e} + {}^3\tx{He}_{\tx{D}+} + e_{\tx{D}+}(\tx{bound})
	&\to {}^3\tx{H}_{\tx{D}-},
    \end{align*}
    where $^3\tx{H}_{\tx{S}+}$, $^3\tx{He}_{\tx{S}-}$ and $e_{\tx{S}-}$ represent incoming tritium, outgoing helium-3 and electron, which form bound states with helium-3, while $^3\tx{He}_{\tx{D}+}$, $e_{\tx{D}+}$ and $^3\tx{H}_{\tx{D}-}$ represent incoming helium-3, electron bounded with helium-3, and outgoing tritium. 

    This indeed has the structure of a $t$-channel--like scattering diagram. For an evaluation of neutrino oscillations in the neutrino M\"ossbauer process within the framework of WPQFT, see for example Ref.~\cite{Akhmedov:2008jn}.
}
    \item The Jacob--Sachs theorem~\cite{Jacob:1961zz} provides the asymptotic form valid when the propagation time $\Delta T$ is sufficiently long. 
    For shorter values of $\Delta T$, how should the asymptotic formula~\eqref{eq: Jacob-Sachs theorem} be modified? 
    If the behavior differs substantially, such effects may become relevant for BSM particles.
\end{itemize}
In this section, to address these questions, we directly evaluate the integral over the zeroth component of the propagation momentum in the amplitude~\eqref{eq: effective propagator} by means of the saddle-point method. 

\subsection{Evaluation of propagating energy integration}
\begin{figure*}[t]
  \centering
  {\includegraphics[width=0.46\textwidth]{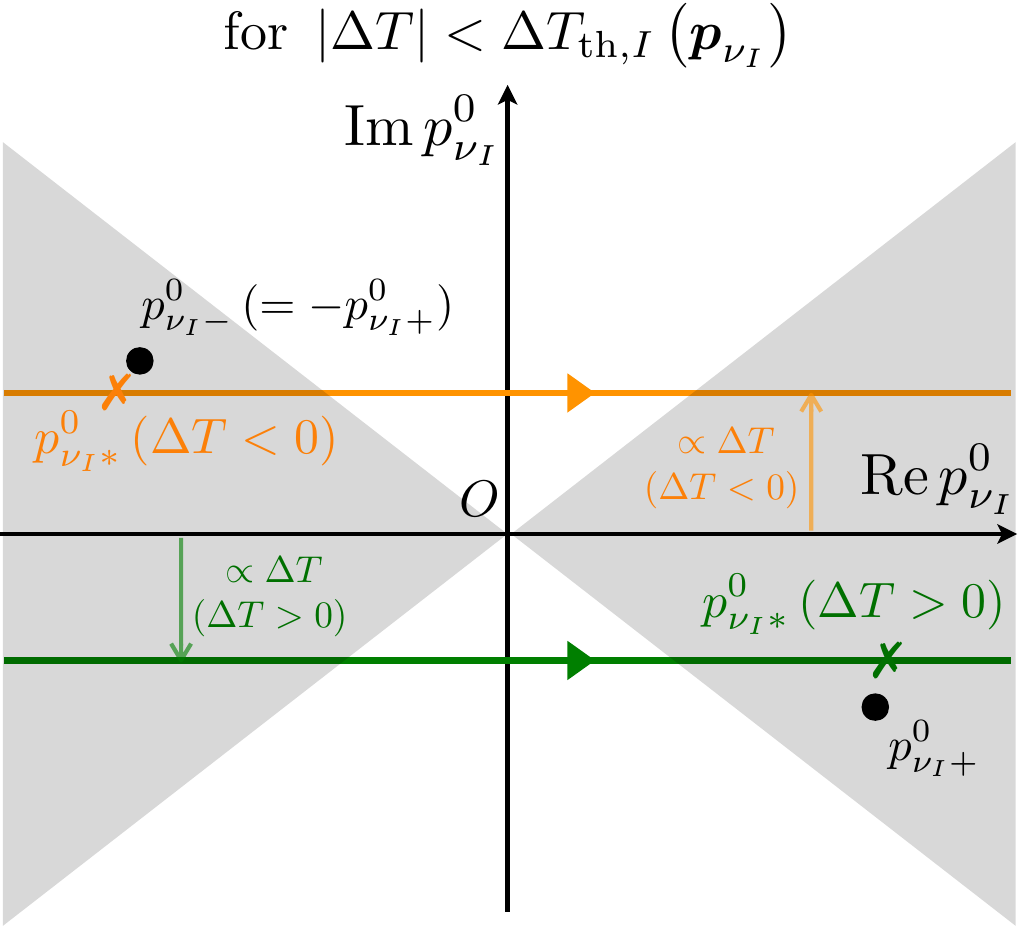}}
  {\includegraphics[width=0.46\textwidth]{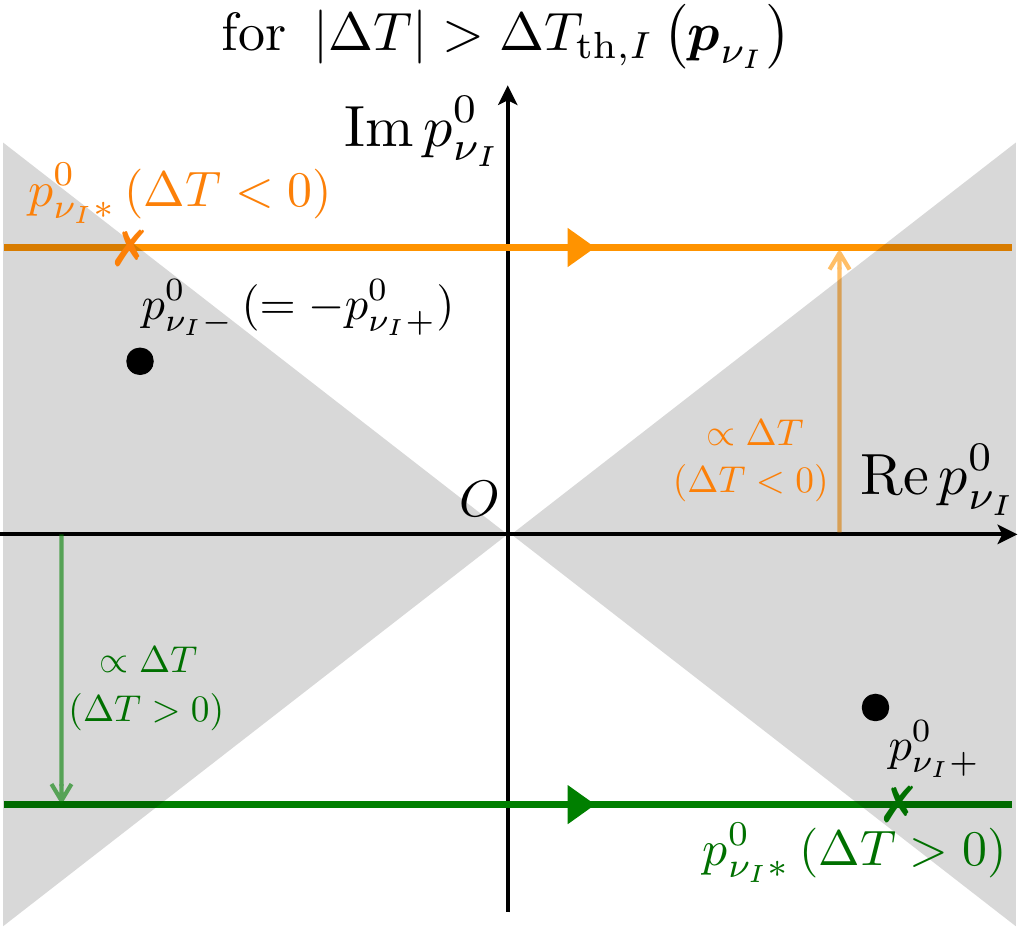}}
  \caption{Schematic integration contours for the energy integral of the propagation amplitude. Colored lines show the contours passing through the saddle points, which are depicted as crosses, and black dots indicate the pole positions. The shaded regions represent the directions where the integrand, including the overlap function, is convergent as $|p_{\nu_I}^0| \to \infty$. For large $\Delta T$, the saddle point has a larger imaginary part. Thus, the contour deformation crosses the propagator pole, yielding a contribution from real-particle propagation.
  }
  \label{fig: Integration contour}
\end{figure*}

We consider directly evaluating the energy integral in Eq.~\eqref{eq: effective propagator} using the saddle-point method with the Gaussian WPQFT framework~\cite{Ishikawa:2018koj, Ishikawa:2020hph, Ishikawa:2021bzf, Ishikawa:2023bnx}. In this approach, the integral can be decomposed into two main contributions. One arises from the saddle point, while the other originates from the poles of the propagator~\cite{Ishikawa:2020hph, Ishikawa:2021bzf}. The latter contribution appears because, when deforming the integration contour from the real axis to the path passing through the saddle point, the contour crosses a pole if the imaginary part of the saddle point is larger than that of the pole. In Fig.~\ref{fig: Integration contour}, we illustrate the integration contour and its relation to the positions of the poles.

The poles of the propagator with respect to $p_{\nu_I}^0$, $p_{\nu_I \pm}^0$, and the saddle point of the energy integral, $p_{\nu_I \star}^0$, are given by
\begin{align}
p_{\nu_I \pm}^0
    &:=
        \pm \sqrt{\bs{p}_{\nu_I}^2 + m_{\nu_I}^2 - i m_{\nu_I} \Gamma_{\nu_I}},\\
p_{\nu_I \star}^0
    &:=
        P^{0}_{\nu}\fn{\bs{p}_{\nu_I}} - i \sigma_{t, \mr{S}+\mr{D}}^{-1} \Delta T.
        \label{eq:pzero-ast}
\end{align}
Here, $P^{0}_{\nu}\fn{\bs{p}_{\nu_I}}$ denotes the real part of the saddle point, 
\begin{align}
P^0_{\nu} \fn{\bs{p}_{\nu_I}}
	&:=
          \sumSandD \pn{-1}^{\SD} { \sigma_{t, \SD} \over  \sigma_{t, \tx{S}+ \tx{D}}}
		 	\ol{P^0}_{\SD 0}\nonumber\\
	&\quad
        + \sumSandD \pn{-1}^{\SD} { \sigma_{t, \SD} \over  \sigma_{t, \tx{S}+ \tx{D}}}
		 	\pn{ \pn{-1}^{\SD} \bs{p}_{\nu_I}  - \ol{\bs{P}}_{\SD 0}} \cdot \ol{\bs{v}}_{\SD 1}.
\end{align}
where $\sum_{\SD=\tx S,\tx D}$ denotes the sum over the contributions from S and D. 
The factor $\pn{-1}^{\SD}$ is a sign determined by the label:
\begin{align}
\pn{-1}^{\SD}
    :=
\begin{cases}
+1 & (\SD=\tx S),\\ 
-1 & (\SD=\tx D).   
\end{cases}
\label{appeq: pm SD notation}
\end{align}

To perform the saddle-point approximation, the original integration contour along the real axis must be deformed to the steepest-descent path passing through the saddle point, as shown in Fig.~\ref{fig: Integration contour}.
If this deformation crosses a pole, the residue of the crossed pole must be added in addition to the saddle-point contribution. The contour crosses a pole during the deformation in the following two cases:
\begin{align}
\begin{cases}
\Delta T > 0 ~\text{and}~ \sigma_{t, \tx{S} + \tx{D}}^{-1} \Delta T 
+ \im[p_{\nu_I +}^0]
> 0, \\[5mm]
\Delta T < 0 ~\text{and}~ -\sigma_{t, \tx{S} + \tx{D}}^{-1} \Delta T 
- \im[p_{\nu_I -}^0]
> 0,
\end{cases}
\end{align}
where $\im[p_{\nu_I +}^0] = - \im[p_{\nu_I -}^0]$.
The first case corresponds to a forward-in-time process ($\Delta T > 0$) in which the pole associated with positive energy is picked up. The second case corresponds to a backward-in-time process ($\Delta T < 0$) in which the pole associated with negative energy is picked up~\cite{Ishikawa:2020hph, Ishikawa:2021bzf}. The backward-in-time contribution is, as will be shown later, incompatible with energy conservation and is therefore highly suppressed compared to the forward-in-time contribution~\cite{Ishikawa:2020hph, Ishikawa:2021bzf}.

This direct evaluation of the amplitude using the saddle-point method resolves one of the questions raised at the beginning of this section. In particular, the pole corresponding to a neutrino propagation that appears ``$t$-channel-like’’ at first sight is picked up through the deformation of the integration contour. The deformation of the contour is controlled by the propagation time $\Delta T$. If the absolute value of $\Delta T$ is larger than the threshold time\footnote{
$\displaystyle {m_{\nu_I} \Gamma_{\nu_I} \over 2 E_{\nu_I}\fn{\bs{p}_{\nu_I}}}
\simeq \left| \im[p_{\nu_I+}^0] \right| = \left| \im[p_{\nu_I-}^0] \right|$.
This approximation, and hence Eq.~\eqref{eq: time threshold approx}, holds for $m_{\nu_I} \Gamma_{\nu_I} \ll E_{\nu_I}^2$, with relative corrections of $O\fn{\pn{m_{\nu_I} \Gamma_{\nu_I}/E_{\nu_I}^2}^2}$.
Note also that if we take the plane wave limit naively $\sigma_{t,\tx{S}+\tx{D}} \to \infty$, $\Delta T_{\mr{th},I}$ goes to infinity, meaning that all of the intermediate neutrinos remain off-shell.
}
\begin{align}
\label{eq: time threshold}
\Delta T_{\mr{th},I} \fn{\bs{p}_{\nu_I}}
    &:= \sigma_{t, \tx{S} + \tx{D}} 
    \sqrt{\sqrt{E_{\nu_I}^{4} + \pn{m_{\nu_I} \Gamma_{\nu_I}}^2}-E_{\nu_I}^{2} \over 2 }\\
    &\simeq 
        \sigma_{t, \tx{S} + \tx{D}} \,
    {m_{\nu_I} \Gamma_{\nu_I} \over 2 E_{\nu_I}\fn{\bs{p}_{\nu_I}}},
    \label{eq: time threshold approx}
\end{align}
the neutrino can become a real particle. In this case, the amplitude exhibits the asymptotic behavior implied by the Jacob--Sachs theorem~\eqref{eq: Jacob-Sachs theorem}. Physically, the threshold time increases with $\Gamma_{\nu_I}$ because a larger decay width displaces the pole further from the real axis in the complex $p_{\nu_I}^0$ plane; consequently, the integration contour must be deformed by a larger amount---i.e., $\Delta T$ must be larger--- to capture the pole. On the other hand, when the propagation time is shorter than the threshold, the neutrino propagation is instead described solely by the saddle-point contribution. As will be seen explicitly in a later section, propagation determined by the saddle-point contribution cannot occur over a classical distance. For this reason, we refer to it as {\it virtual} neutrino propagation.

We note that the $t$-channel singularity discussed in the scattering of (un)stable particles~\cite{Peierls:1961zz, Coleman:1965xm, Nowakowski:1993iu, Brayshaw:1978xt, Ginzburg:1995bc, Melnikov:1996na, Melnikov:1996iu} can appear in the saddle-point contribution. Therefore, a virtual neutrino can also hit the pole for certain configurations of the external momenta. However, this contribution does not appear in the asymptotic form for long propagation times~\eqref{eq: Jacob-Sachs theorem}, and it arises only when the configuration of the external momenta is highly tuned.

In the saddle-point contribution, the flavor-changing effect appears only through the mass dependence contained in $M_{I,\alpha\beta}\fn{m_{\nu_I},\, p^0_{\nu_I},\, \bs{p}_{\nu_I}}$. Hence, when the neutrino is relativistic, the corresponding transition probability is strongly suppressed.

\begin{figure*}[t]
\centering
\begin{tikzpicture}

\node (left) at (0,0) {
\begin{tikzpicture}
\begin{feynman}
  \vertex (vS) at (0,0);
  \vertex [left=1cm of vS] (nS) ;
  \vertex [right=1.5cm of vS] (pS) ;
  \vertex [above right=1cm of vS] (eS) 
  {\(l_{\alpha}\)};
  \vertex (vD) at (0.5,-2);
  \vertex [left=1.5cm of vD] (pD) ;
  \vertex [below right=1cm of vD] (eD)
   {\(l_{\beta}\)};
  \vertex [right=1cm of vD] (nD) ;

  \diagram* {
    (nS) -- (vS),
    (vS) -- (pS),
    (vS) -- (eS),
    (vS) -- [edge label=\(\nu_\alpha \to \nu_\beta\)] (vD),
    (pD)--  (vD),
    (eD)--(vD),
    (vD)--(nD),
  };
\end{feynman}
\end{tikzpicture}
};

\node at (2,0) {\Large $\simeq$};

\node (left) at (4,0) {
\begin{tikzpicture}
\begin{feynman}

  \vertex (vS) at (0,0);
  \vertex [left=1cm of vS] (nS);
  \vertex [right=1.5cm of vS] (pS);
  \vertex [above right=1cm of vS] (eS)
  {\(l_{\alpha}\)};

  \vertex (vD) at (0.5,-2);
  \vertex [left=1.5cm of vD] (pD);
  \vertex [below right=1cm of vD] (eD)
  {\(l_{\beta}\)};
  \vertex [right=1cm of vD] (nD);

  \vertex at ($(vS)!0.5!(vD)$) (vX) ;

  \diagram* {
    (nS) -- (vS),
    (vS) -- (pS),
    (vS) -- (eS),

    (vS) --[edge label=\(\nu_\alpha\)] (vX)
          --[edge label=\(\nu_\beta\)] (vD),

    (pD)-- (vD),
    (eD)--(vD),
    (vD)--(nD),
  };

  \node(cross)  at (vX) {\Large $\times$};
\end{feynman}
\end{tikzpicture}
};

\node at (7.5,0) {{\Large $+$} {\large$\sum_I \theta\fn{\Delta T - \Delta T_{\mr{th,I}}} \times$}};

\node (right) at (11,0) {
\begin{tikzpicture}
\begin{feynman}
  \vertex (vS1) at (0,0);
  \vertex [left=1cm of vS] (nS) ;
  \vertex [right=1.5cm of vS] (pS) ;
  \vertex [above right=1cm of vS] (eS) 
   {\(l_{\alpha}\)};
  \vertex (vS2) at (0.25,-1) {\(\nu_I\)};
  \vertex (vD1) at (0.25,-2) {\(\nu_I\)};
  \vertex (vD2) at (0.5,-3);
  \vertex [left=1.5cm of vD2] (pD) ;
  \vertex [below right=1cm of vD2] (eD) 
  {\(l_{\beta}\)};
  \vertex [right=1cm of vD2] (nD) ;

  \diagram* {
    (nS) -- (vS1),
    (vS) -- (pS),
    (vS) -- (eS),
    (vS1) -- (vS2),
    (vD1) -- (vD2),
    (pD)-- (vD2),
    (eD)--(vD2),
    (vD2)--(nD),
  };
\end{feynman}
\end{tikzpicture}
};
\end{tikzpicture}
\caption{Feynman-diagrammatic interpretation of the amplitude evaluation using the saddle-point method. The first diagram on the right-hand side corresponds to the contribution from the saddle point and can be interpreted as the propagation of a virtual particle. Flavor transition occurs through a mass insertion (indicated by a cross in the figure). The second diagram represents the contribution that arises when the integration contour crosses a pole of the propagator ($\theta(x)$ denotes the step function). This can be interpreted as the propagation of a real particle.}

\label{fig: extended Jacob-Sachs theorem}
\end{figure*}
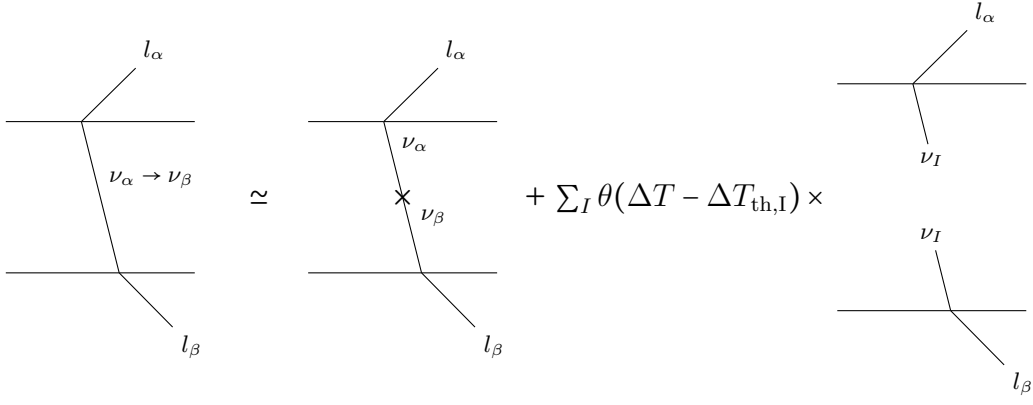

We present in Fig.~\ref{fig: extended Jacob-Sachs theorem} a Feynman-diagrammatic interpretation of the two-flavor transition amplitudes that arise from the saddle-point method.

In summary, the diagram describing the flavor transition has two contributions, and one of them dominates depending on the propagation time. When the propagation time is shorter than the threshold~\eqref{eq: time threshold}, the transition is induced by flavor flipping associated with the mass-inserted term. When the propagation time exceeds the threshold~\eqref{eq: time threshold}, the flavor transition occurs due to flavor oscillation arising from the mass dependence in the exponential.

\subsection{Amplitude and Its Plane Wave Limit}
\label{subsec: Amplitude and Plane wave limit}
After performing the energy integration, the momentum integration can also be evaluated using the saddle-point method in the same manner. Therefore, deformation of the integration contour is again required; however, no additional contributions arise from poles. From the above considerations, it follows that the expression for the amplitude can be classified into the following two cases: amplitude for the real neutrinos and virtual neutrinos. The derivation of the expression for the amplitude presented in this subsection is deferred to Appendix~\ref{app: Evaluation of the internal-momentum integration}.

First, the real propagating contribution to the amplitude, $\mc A^{\mr{re}}_{I, \alpha\beta}$, appears when the condition $\Delta T > \Delta T_{\mr{th},I}\fn{\bs{P}_{\nu}}$ is satisfied. Here, 
\begin{align}
\bs{P}_{\nu}
	&:=
	\sumSandD \pn{-1}^{\SD} {\sigma_{\SD} \over \sigma_{\tx{S}+\tx{D}}}\,\ol{\bs{P}}_{\SD0}
    \label{eq:P-spatial_nu}
\end{align}
denotes the real part of the saddle point of the propagation momentum. The expression for $\mc A^{\mr{re}}_{I, \alpha\beta}$ can be written in the following form:
\begin{align}
\label{eq: amplitude on-shell}
\mc A^{\mr{re}}_{I, \alpha\beta}
    &\simeq
    N_{X} \psi_{X} 
			 \ \pn{2 \pi}^4 \pn{\sigma \over 2 \pi}^{3/2} \pn{\sigma_{t} \over 2 \pi}^{1/2} \psi_P\nonumber\\
	&\quad
		\times 
		 \Bigg[
				 {
				 	 M_{I,\alpha\beta}
                        \fn{m_{\nu_I}, \, \sqrt{z_{\nu_I} + \bs{P}_{\nu}^2},\bs{P}_{\nu}}
                    \over  
                    2 \sqrt{z_{\nu_I} + \bs{P}_{\nu}^2}}
				\, e^{i \varphi} \nonumber\\
    &\phantom{\quad\times\Bigg[}
		\times
			{2 \pi} \pn{\sigma_{t, \tx{S} + \tx{D}} \over 2 \pi }^{1/2} 
				\psi_{\tx{on}}\fn{E_{\nu_I} \fn{\bs{P}_{\nu}}}\nonumber\\
	&\phantom{\quad\times\Bigg[}
		\times
			\exp\pn{
				- i \sqrt{z_{\nu_I} + \bs{P}_{\nu}^2} \Delta T + i P_{\nu}^i L^i
				}\nonumber\\
    &\phantom{\quad\times\Bigg[}
		\times
			\exp\pn{
					- {1 \over 2\sigma_{\tx{S}+\tx{D}}}
							\pn{L^i - v_{\nu_I}^i \fn{\bs{P}_{\nu}} \Delta T}^2
					}
					\Bigg],
\end{align}
where
\begin{align}
N_X 
    &:= \pn{
			\prod_{\SD=\tx{S}, \tx{D}}
            \prod_{\mc{N}_{\tx{R}\pm}}
			{\pn{\pi \sigma_{\mc{N}_{\SD \pm}}}^{- {3 \over4}}
			\over
			\sqrt{2 E_{\mc{N}_{\SD}}\fn{\bs{P}_{\mc{N}_{\SD\pm}}}} 
			}
			},\\
\sigma_{t} 
	&:=
        {\sigma_{t, \tx{S}} \sigma_{t, \tx{D}} \over \sigma_{t, \tx{S} +\tx{D}}}, \\
\label{eq: sigma}
\sigma 
	&:=
        {\sigma_{\tx{S}} \sigma_{\tx{D}} \over \sigma_{\tx{S} +\tx{D}}}.
\end{align}
Here, we recall Eq.~\eqref{eq:z_nu_I-definition} for $z_{\nu_I}$.
$\psi_{X}$ and $\psi_{P}$ characterize the spatial localization of the external legs at the interaction points and the momentum localization that ensures energy-momentum conservation of the external legs, while $\psi_{\tx{on}}$ describes the on-shell resonance:
\begin{align}
\label{eq: psiX localization}
\psi_{X}
	&:=	\exp\pn{
				- {1\over2} \sumSandD \sigma_{\SD}^{-1}
				\pn{
					\ol{X^i_{\medstar} \fn{T_{\SD}}^2}_{\SD1} 
					-  \pn{X^i_{\SD}}^2
					}},\\
\label{eq: psiP localization}
\psi_{P}
	&:=
		 \exp\pn{
		-{\sigma \over 2}  \pn{\ol{P^i}_{\tx{S}0}+ \ol{P^i}_{\tx{D}0}}^2
		} \nonumber\\
	&\quad\times
        \exp\pn{- {\sigma_{t} \over 2}  \pn{\ol{P^0}_{\tx{S}0} + \ol{P^0}_{\tx{D}0} }^2
				}, \\
\label{eq: psion localization}
\psi_{\tx{on}}\fn{E_{\nu_I}}
	&:=
		\exp\pn{- {\sigma_{t, \tx{S}+\tx{D}} \over 2}  
			\pn{E_{\nu_I}\fn{\bs{P}_{\nu}} - P_{\nu}^0\fn{\bs{P}_{\nu}} }^2
				}.
\end{align}

As will be discussed later, $\psi_{P}$ reduces to the energy--momentum conserving delta function in the plane-wave limit, while $\psi_{\tx{on}}$ reduces to a Breit--Wigner-type delta function. The expression shown in Eq.~\eqref{eq: amplitude on-shell} corresponds to the forward-in-time process ($\Delta T > 0$). For the backward-in-time process ($\Delta T < 0$), the sign of $E_{\nu_I}\fn{\bs{P}_{\nu}}$ appearing in the exponent of $\psi_{\tx{on}}\fn{E_{\nu_I}}$ is reversed. For the configurations of interest, in which the external particles lose energy in the production process and gain energy in the detection process, the real part of the saddle point of the propagation energy, $P_{\nu}^0\fn{\bs{P}_{\nu}}$, is positive. Hence, there is no configuration for which the exponent vanishes~\cite{Ishikawa:2020hph, Ishikawa:2021bzf}, and the corresponding amplitude is strongly suppressed compared with the forward-in-time process.

In addition, the amplitude~\eqref{eq: amplitude on-shell} contains, in the exponent, a mass-dependent phase,
\begin{align}
\label{eq: mass dependent phase}
&e^{i\sqrt{z_{\nu_I} + \bs{P}_{\nu}^2} \Delta T}
        =e^{i \sqrt{m_{\nu_I}^2 + \bs{P}_{\nu}^2 - i m_{\nu_I} \Gamma_{\nu_I}} \, \Delta T}.
\end{align}
This phase is equivalent to that obtained from the Jacob--Sachs theorem~\eqref{eq: Jacob-Sachs theorem}, and, when converted into a probability, it leads to the flavor-oscillation behavior governed by the mass-squared differences.

Finally, the last factor in Eq.~\eqref{eq: amplitude on-shell} plays the role of suppressing configurations that deviate from the classical trajectory,
\begin{align}
\label{eq: classical propagation}
L^i - v_{\nu_I}^i \fn{\bs{P}_{\nu}} \Delta T \simeq 0 .
\end{align}
This behavior is consistent with previous studies~\cite{Giunti:1997wq, Beuthe:2001rc, Giunti:2002xg,  Akhmedov:2010ms}.

On the other hand, the virtual propagating contribution, $\mc A^{\mr{vi}}_{I, \alpha\beta}$, corresponds to the saddle-point contribution and is given by
\begin{align}
\label{eq: amplitude off-shell}
\mc A^{\mr{vi}}_{I, \alpha\beta}
	&\simeq
			N_X \psi_{X} 
		\
			 \pn{2 \pi}^4 \pn{\sigma \over 2 \pi}^{3/2} \pn{\sigma_{t} \over 2 \pi}^{1/2} \psi_P \nonumber\\
	&
		\quad \times 
				{(-i) \, M_{I,\alpha\beta}
                        \fn{m_{\nu_I}, \, P^0_{\nu}\fn{\bs{P}_{\nu}},
                        \bs{P}_{\nu}}\, 
                        e^{i \varphi}
			\over 
				{- P^0_{\nu}\fn{\bs{P}_{\nu}}^2 
                + \bs{P}_{\nu}^2 + m_{\nu_I}^2 - i m_{\nu_I} \Gamma_{\nu_I}}} \nonumber\\
	&\quad
		\times 
            \exp\pn{
		-{1 \over 2\sigma_{\tx{S}+ \tx{D}}}  \pn{ L^i -  v^i_{\nu, \tx{vi}} \Delta T}^2
		}\nonumber\\
	&\quad
		\times
		\exp\pn{- {1\over2\sigma_{t, \tx{S} + \tx{D}}} \Delta T^2},
\end{align}
where
\begin{align}
\label{eq: voff}
v^i_{\nu, \text{vi}}
	&:=
        {\sigma_{t, \text{S}} \ol{v^i}_{\text{S} 1} +\sigma_{t, \text{D}} \ol{v^i}_{\text{D} 1}\over  \sigma_{t, \tx{S}+ \tx{D}}}.
\end{align}

The characteristic features of this contribution are the absence of a mass-dependent phase, such as the one appearing in Eq.~\eqref{eq: mass dependent phase}, and the presence of a suppression factor that favors short propagation times,
\begin{align}
\label{eq: short time propagation}
\exp\pn{- {1 \over 2\sigma_{t, \tx{S} + \tx{D}}} \Delta T^2} .
\end{align}
The difference in the phase structure implies that flavor oscillations do not occur in the virtual regime.
In addition, in typical long-baseline neutrino oscillation experiments, the effects of virtual propagation cannot be directly observed due to this suppression factor~\eqref{eq: short time propagation}.\footnote{However, as already mentioned in the previous section, the pole in the denominator of Eq.~\eqref{eq: amplitude off-shell} may be hit for certain parameter configurations. This effect becomes important, for example, for neutrinos with short propagation times that appear in collider contexts~\cite{Brayshaw:1978xt, Ginzburg:1995bc, Melnikov:1996na, Melnikov:1996iu}.}

These amplitude expressions provide an answer to the second question raised at the beginning of this section. In particular, when the propagation time becomes shorter than a certain threshold, $\Delta T < \Delta T_{\mr{th},I}\fn{\bs{P}_{\nu}}$, the real propagating contribution corresponding to the asymptotic form suggested by the Jacob--Sachs theorem~\eqref{eq: Jacob-Sachs theorem} disappears. In this regime, the amplitude is dominated solely by the contribution from virtual propagation arising from the saddle point, which is described by Eq.~\eqref{eq: amplitude off-shell} as $\mc A^{\mr{vi}}_{I, \alpha\beta}$.

Finally, at the end of this subsection, we examine the behavior of each amplitude $\mc A^{\mr{re}(\mr{vi})}_{I, \alpha\beta}$ in the plane-wave limit. In the plane-wave limit, the momentum localization factor $\psi_P$ becomes a delta function corresponding to the energy-momentum conservation, while $\psi_{\tx{on}}$ becomes a delta function which describes the on-shell pole,
\begin{align}
\label{eq: energy momentum conservation in plane wave limit}
\pn{\sigma \over 2 \pi}^{3/2} \pn{\sigma_{t} \over 2 \pi}^{1/2} \psi_P
	&\underset{\sigma \to \infty}{\longrightarrow} 
		\delta^4 \fn{\ol{P}_{\tx{S}0} + \ol{P}_{\tx{D}0}}, \\
\pn{\sigma_{t, \tx{S} + \tx{D}} \over 2 \pi }^{1/2} 
				\psi_{\tx{on}}\fn{E_{\nu_I} \fn{\bs{P}_{\nu}}}
	&\underset{\sigma \to \infty}{\longrightarrow} 
        \delta \fn{E_{\nu_I}\fn{\bs{P}_{\nu}} - P^0_{\nu}\fn{\bs{P}_{\nu}}},	
\end{align}
where we refer to Eq.~\eqref{eq:pmC-operation} for the relative signs of the arguments.

Accordingly, the amplitudes corresponding to real and virtual propagation, $\mc A^{\mr{re}(\mr{vi})}_{I,\alpha\beta}$, take the following forms in the plane-wave limit:
\begin{align}
\label{eq: amplitude on-shell PW limit}
\mc A^{\mr{re}}_{I, \alpha\beta}
    &\underset{\sigma \to \infty}{\longrightarrow} 
        N_{X} \psi_{X} \ \pn{2 \pi}^4 \delta^4 \fn{\ol{P}^{\mu}_{\tx{S}0} + \ol{P}^{\mu}_{\tx{D}0}}\nonumber\\
    &\qquad
        \times {2 \pi} \delta \fn{E_{\nu_I}\fn{\bs{P}_{\nu}} - P^0_{\nu}\fn{\bs{P}_{\nu}}}\nonumber\\
	&\qquad
		\times 
		 \Bigg[
				 {
				 	 M_{I,\alpha\beta}
                        \fn{m_{\nu_I}, \, \sqrt{z_{\nu_I} + \bs{P}_{\nu}^2},\bs{P}_{\nu}}
                    \over  
                    2 \sqrt{z_{\nu_I} + \bs{P}_{\nu}^2}}
				\, e^{i \varphi} \nonumber\\
	&\phantom{\qquad\times\Bigg[}
			\exp\pn{
				- i \sqrt{z_{\nu_I} + \bs{P}_{\nu}^2} \Delta T + i P_{\nu}^i L^i
				}
					\Bigg], \\
\mc A^{\mr{vi}}_{I, \alpha\beta}
	&\underset{\sigma \to \infty}{\longrightarrow} 
			N_{X} \psi_{X}  \ \pn{2 \pi}^4 \delta^4 \fn{\ol{P}_{\tx{S}0} + \ol{P}_{\tx{D}0}} \nonumber\\
	&
		\qquad \times 
				{M_{I,\alpha\beta}
                        \fn{m_{\nu_I}, \, P^0_{\nu}\fn{\bs{P}_{\nu_I}},
                        \bs{P}_{\nu}}\, 
                        e^{i \varphi}
			\over 
				{- P^0_{\nu}\fn{\bs{P}_{\nu_I}}^2 
                + \bs{P_{\nu}}^2 + m_{\nu_I}^2 - i m_{\nu_I} \Gamma_{\nu_I}}}.
\end{align}

Interestingly, the plane-wave limit of $\mc A^{\mr{re}}_{I,\alpha\beta}$ takes a form as if the propagator were approximated by a Breit--Wigner-type delta function. 
This suggests an effective cutting rule.

Namely, in the limit $\Delta T \to \infty$, the full process including neutrino propagation can be effectively factorized into two subprocesses: the production process [see Eq.~\eqref{eq: interaction at source region}] and the detection process [see Eq.~\eqref{eq: interaction at detection region}]. 
In addition, in the plane-wave limit, the energy is uniquely fixed to the saddle-point value, and therefore, no transitions occur between different flavors.

Apart from the overall factor $N_X \psi_X$, the plane-wave limit of the amplitude for virtual propagation, $\mc A^{\mr{vi}}_{I,\alpha\beta}$, reduces to the familiar form obtained in the standard plane-wave calculation. 
This shows that the present WPQFT framework provides a generalization of the conventional plane-wave approach.

\section{Flavor Changing Probability}
\label{sec: Flavor Changing Probability}
In the previous Sec.~\ref{sec: Real and Virtual Propagation of intermediate particles}, we showed that the flavor-transition amplitude obtained by the saddle-point method is separated into two contributions: the virtual-propagation contribution $\mc A^{\mr{vi}}_{I,\alpha\beta}$ and the real-propagation contribution $\mc A^{\mr{re}}_{I,\alpha\beta}$. Their explicit forms are given in Eqs.~\eqref{eq: amplitude off-shell} and \eqref{eq: amplitude on-shell}, respectively.
The total amplitude can therefore be written as
\begin{align}
\label{eq: flavor changing amplitude}
\mc A_{I,\alpha\beta} 
    = \mc A^{\mr{vi}}_{I,\alpha\beta} 
    + \theta\fn{\Delta T - \Delta T_{\mr{th},I}\fn{\bs{P}_{\nu}}} 
      \mc A^{\mr{re}}_{I,\alpha\beta}.
\end{align}
The flavor-transition probability is obtained by taking the squared amplitude and integrating it over the propagation time, as in Eq.~\eqref{eq: probability given by time integration}.

\subsection{Stepwise structure of the probability}
\label{subsec: Stepwise structure of the flavor-transition probability}
When evaluating the squared amplitude, the interference term between the virtual- and real-propagation contributions,
$\mc A^{\mr{vi}}_{I,\alpha\beta}(\mc A^{\mr{re}}_{J,\alpha\beta})^{*}$, can be neglected. The reason is that these two contributions have support in parametrically different regions of $\Delta T$.
Both the real and virtual amplitudes have unsuppressed configurations at $\Delta T \sim L/\left| v_{\nu_I} \right|$ and $\Delta T \sim L/\left| v_{\nu,\tx{vi}} \right|$, respectively, while the virtual-propagation contribution has an extra exponential suppression away from $\Delta T=0$, whereas the real-propagation contribution appears only after the threshold condition $\Delta T>\Delta T_{\mr{th},I}$ is satisfied.
Therefore, except for exceptional configurations, e.g., where $\Delta T$ is slightly larger than $\Delta T_{\text{th},I}$ and is also close to zero, the interference terms can be safely neglected.

Thus, below the threshold, the probability is controlled by virtual propagation, while above the threshold, it is controlled by real propagation. For relativistic particles, however, the flavor-changing effect induced by virtual propagation is negligible because it arises only through the mass dependence of the non-oscillatory part of the amplitude. Consequently, observable flavor oscillations start only after the relevant threshold time is exceeded within the step-function approximation in Eq.~\eqref{eq: flavor changing amplitude}. The smoothing of this threshold and the resulting onset time scale are discussed in Sec.~\ref{subsec: Smoothing of the propagating time threshold}.

Therefore, when considering flavor transitions only among the SM neutrinos, the flavor-transition probability in the case of normal ordering $m_{\nu_1}<m_{\nu_2}<m_{\nu_3}$ can be expressed as
\begin{align}
\label{eq: stepwise probability}
&P(\alpha \to \beta)\nonumber \\
    &
        \simeq
\begin{cases}
\mc{O}\fn{\sum_I {m_{\nu_I}^2 \over E_{\nu_I}^2}} \ll 1, 
    & \text{for} ~\Delta T_0 < \Delta T_{\mathrm{th},2}\fn{\bs{P}_{\nu}}, \\
P_{12}(\alpha \to \beta)
    & \text{for} ~
    \Delta T_{\mathrm{th},2}\fn{\bs{P}_{\nu}}
    < \Delta T_0 < \Delta T_{\mathrm{th},3}\fn{\bs{P}_{\nu}},\\
P_{\mr{full}}(\alpha \to \beta)
    & \text{for} ~
    \Delta T_{\mathrm{th},3}\fn{\bs{P}_{\nu}}
    < \Delta T_0,
\end{cases}
\end{align}
where
\begin{align}
\label{eq: 12 oscillation}
P_{12}(\alpha \to \beta)
    &\propto
       \sum_{I,J = 1,2}
       {
        M_{I,\alpha\beta} M^*_{J,\alpha\beta}
       \over
       \sum_{K,L=1}^{3} M_{K,\alpha\alpha} M^*_{L,\alpha\alpha}
       }
       \nonumber\\
     &\qquad
     	\times\exp\fn{
				- {L \over L^{\mr{dec}}_{IJ}}
				-2\pi i\, \frac{L}{L^{\mathrm{osc}}_{IJ}}
				-\left(\frac{L}{L^{\mathrm{coh}}_{IJ}}\right)^{2}
				},\\
\label{eq: full oscillation}
P_{\mr{full}}(\alpha \to \beta)
    &\propto
       \sum_{I,J=1}^{3}
       {
       M_{I,\alpha\beta} M^*_{J,\alpha\beta}
       \over
       \sum_{K,L=1}^{3} M_{K,\alpha\alpha} M^*_{L,\alpha\alpha}
       }
       \nonumber\\
     &\qquad
            \times\exp\fn{
				- {L \over L^{\mr{dec}}_{IJ}}
				-2\pi i\, \frac{L}{L^{\mathrm{osc}}_{IJ}}
				-\left(\frac{L}{L^{\mathrm{coh}}_{IJ}}\right)^{2}
				},
\end{align}
with
\begin{align}
\Delta T_{0}
		& := {L^i \bar{v}_{0}^i
			\over 
            \pn{\bar{v}_{0}^i}^2 }, \\  
\ol{v}_{0}^i 
	&:= {P^i_{\nu} \over |\bs{P}_{\nu}|}.
\end{align}
We recall Eqs.~\eqref{eq:def_Losc}, \eqref{eq:def_Lcoh}, and \eqref{eq:def_Ldec} for $L^{\mathrm{osc}}_{IJ}$, $L^{\mathrm{coh}}_{IJ}$, and $L^{\mathrm{dec}}_{IJ}$.
Here, again, we omit the exponential terms that are independent of $L$.
We provide the derivation of these expressions in the Appendix~\ref{app: Evaluation of the propagating time integration}.

We note that when the propagation distance $L$ is classically large, $\psi_X \simeq 1$, or equivalently,
\begin{align}
{
					\ol{X^i_{\medstar} \fn{T_{\SD}}^2}_{\SD1} 
					-  \pn{X^i_{\SD}}^2
					}
	\simeq0,
    \label{eq:X-approximation-condition}
\end{align}
provides a good approximation. The approximation $\psi_X \simeq 1$ expresses the situation in which the saddle-point trajectories of the wave packets intersect at a small interaction region because the exponent of $\psi_X$ represents the uncertainty in the position of the interaction point.
Thus, the condition $\psi_X \simeq 1$ is naturally justified when the propagation distance scale $L$ is sufficiently larger than the uncertainty scale. In this regime, one may neglect the interaction-time dependence through $T_{\SD}$ in $\psi_X$ when performing the propagation-time integral.\footnote{
As is evident from Eqs.~\eqref{eq:def_Xstar}, \eqref{eq: XSD}, and \eqref{eq:def_sigma_tSD}, $T_{\SD}$ and $X^i_{\SD}$ are distinct functions of the external-line profiles.
Therefore, strictly speaking, manually imposing the condition~\eqref{eq:X-approximation-condition} gives rise to a mathematical contradiction.
Here, we phenomenologically treat $T_{\SD}$ and $X^i_{\SD}$ as independent degrees of freedom; subsequently, after imposing the conditions of Eq.~\eqref{eq:X-approximation-condition}, we incorporate only the integral over the degrees of freedom of $\Delta T \, \fn{= T_{\tx{D}} - T_{\tx{S}}}$ (which remain independent even after this imposition) as the degrees of freedom of the external lines in phase space.
(This procedure is also consistent with the physical premise that transition times are not directly measured.)
Of course, within the scope of this phenomenological treatment, nothing can be said regarding the normalization of the overall reaction; however, this aspect will not be discussed in the present paper, but rather left as a subject for our future works.
\label{footnote:Delta-T-integral}
}

\subsection{Upper limit of propagating time threshold}
One may ask whether the threshold can be observed in actual SM neutrino oscillation experiments. Unfortunately, for SM neutrinos, the answer is negative. The threshold propagation time $\Delta T_{\mr{th},I}$ is bounded from above, and this upper bound is far too short to be experimentally accessible.

To make the threshold propagation time $\Delta T_{\mr{th},I} \fn{\bs{P}_{\nu}}$ sufficiently large, the energy uncertainty $\sigma_{t, \tx{S} + \tx{D}}^{-1/2}$ must be made small. However, if this uncertainty is too small, different mass eigenstates become distinguishable.\footnote{
Here, we implicitly assume
\begin{align*}
\sigma_{t, \tx{S} + \tx{D}}^{-1/2} \simeq \sigma_{\tx{S} + \tx{D}}^{-1/2},
\end{align*}
which is satisfied in typical neutrino oscillation experiments. This means that the uncertainties in energy and momentum are of the same order. On the other hand, in the neutrino M\"ossbauer process~\cite{Visscher:1959kka, Kells:1982rm, Kells:1983iac}, discussed in a later paragraph, there exists a hierarchy between the two,
$\sigma_{t, \tx{S} + \tx{D}}^{-1/2} \ll \sigma_{\tx{S} + \tx{D}}^{-1/2}$.
In this case, a small energy uncertainty does not necessarily imply the disappearance of oscillation phenomena~\cite{Akhmedov:2008jn}.
} In that case, oscillations are no longer realized even at long baselines, in conflict with observations. This distinguishability effect is encoded in the exponent of $\psi_{\tx{on}}\fn{E_{\nu_I}}$ in Eq.~\eqref{eq: psion localization}, and the following condition must be satisfied for the normal hierarchy:
\begin{align}
\sqrt{\sigma_{t, \tx{S}+\tx{D}}} \lesssim \frac{ |\bs{P}_{\nu}| }{\Delta m_{13}^2} ,
\end{align}
where we have assumed that $\sigma_{t, \tx{S}+\tx{D}} \simeq \sigma_{\tx{S}+\tx{D}}$.
Therefore, $\Delta T_{\mr{th},3} \fn{\bs{P}_{\nu}}$ has the following upper bound:
\begin{align}
\Delta T_{\mr{th},3} \fn{\bs{P}_{\nu}}
    &\simeq
        \sigma_{t, \tx{S} + \tx{D}} 
    {m_{\nu_3} \Gamma_{\nu_3} \over 2 |\bs{P}_{\nu}|}, \nonumber\\
    &\lesssim
        \pn{|\bs{P}_{\nu}|\over \Delta m_{13}^2}^2
        {m_{\nu_3} \Gamma_{\nu_3} \over |\bs{P}_{\nu}|} \nonumber\\
    &\simeq
       {m_{\nu_3} \Gamma_{\nu_3} \over 4 \pi \Delta m_{13}^2}L^{\mathrm{osc}}_{13}.
\end{align}
Thus, for SM neutrinos, $\Delta T_{\mr{th},3} \fn{\bs{P}_{\nu}}$ cannot reach a scale that is experimentally accessible due to the suppression $m_{\nu_3} \Gamma_{\nu_3} / \Delta m_{13}^2 \ll 1$. 

However, we note that this limitation may be overcome in setups with extremely high energy resolution. For example, the neutrino M\"ossbauer process~\cite{Visscher:1959kka, Kells:1982rm, Kells:1983iac}, being a resonant process, may realize a very small energy uncertainty $\sigma_{t,\tx{S}+\tx{D}}^{-1/2}$.
It would be interesting to study whether the neutrino M\"ossbauer process satisfies the assumptions of the Jacob--Sachs theorem, a question we leave for future work.
For particles whose $\Delta T$ threshold is longer than that of SM neutrinos, establishing a method to precisely measure the behavior of short-distance propagation might also be useful for probing this transition.

In any case, our results clarify the range of applicability of the Jacob--Sachs theorem~\cite{Jacob:1961zz}, which is known as an asymptotic formula, and establish its validity at short propagation distances, a point that had not been clear in the previous literature.\footnote{
As a discussion of a different aspect of short-distance propagation, there are also studies~\cite{Ioannisian:1998ch, Karamitros:2022nnh} that describe the limitation of applying the Grimus--Stockinger theorem~\cite{Grimus:1996av}. They show that the angular-integral expression in Ref.~\cite{Grimus:1996av} ceases to be valid at short distances. On the other hand, our discussion of the threshold of the integral concerns the integral over the radial degree of freedom of the momentum, $|\bs{p}|$, in the terminology of the Grimus--Stockinger theorem. We also note that, under the no-dispersion assumption adopted for simplicity in our analysis, the spherical-wave-like behavior characteristic of the Grimus--Stockinger theorem does not appear, in accordance with the above-mentioned studies~\cite{Ioannisian:1998ch, Karamitros:2022nnh}.
}

\subsection{Smoothing of the propagating time threshold}
\label{subsec: Smoothing of the propagating time threshold}
\begin{figure}[!t]
\centering
\includegraphics[width=\columnwidth]{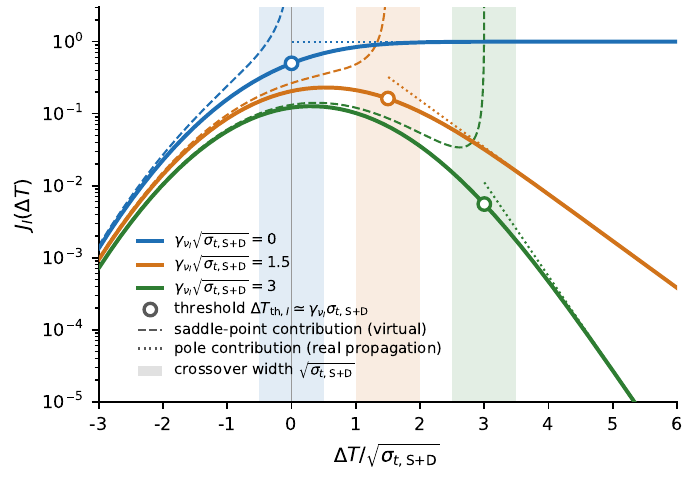}
\caption{
$J_I\fn{\Delta T}$ in Eq.~\eqref{eq: erfc smoothing} for $E_{\nu_I} = P^0_{\nu}$, in which case it is real and positive and depends only on $\Delta T/\sqrt{\sigma_{t, \tx{S}+\tx{D}}}$ and $\gamma_{\nu_I} \sqrt{\sigma_{t, \tx{S}+\tx{D}}}$. The three curves correspond to $\gamma_{\nu_I} \sqrt{\sigma_{t, \tx{S}+\tx{D}}} = 0$, $1.5$, and $3$, from left to right.
Circles mark the thresholds $\Delta T_{\mr{th},I} \simeq \gamma_{\nu_I} \sigma_{t, \tx{S}+\tx{D}}$ in Eq.~\eqref{eq: time threshold approx}, and shaded bands the crossover regions of width $\sqrt{\sigma_{t, \tx{S}+\tx{D}}}$ centered at them.
Dashed lines show the saddle-point contribution, which diverges at the threshold (see the text). Dotted lines show the pole contribution.
\label{fig: erfc crossover}
For nonzero $\gamma_{\nu_I}$, the curves are exponentially suppressed at large $\Delta T$ due to the decay of the propagating particle, as discussed in the text.
}
\end{figure}

The step function in Eq.~\eqref{eq: flavor changing amplitude} is an artifact of the saddle-point method and is smoothed by the exact energy integration.\footnote{
See Ref.~\cite{Ishikawa:2021bzf} for the smoothing of a similar discontinuity.
}
Near the positive-energy pole, the $p^0_{\nu_I}$ integration in the amplitude~\eqref{eq: effective propagator}, with the exponent evaluated in Appendix~\ref{app: Evaluation of the internal-momentum integration}, is captured by the following integral, plotted in Fig.~\ref{fig: erfc crossover}:
\begin{align}
\label{eq: erfc smoothing}
J_I\fn{\Delta T}
&:=	{i\over2\pi}\int_{-\infty}^{\infty} d p_{\nu_I}^0\,
e^{-{\sigma_{t, \tx{S}+\tx{D}}\over2} \pn{p_{\nu_I}^0-P^0_{\nu}}^2}\,
{e^{-i \pn{p_{\nu_I}^0 - P^0_{\nu}} \Delta T} \over p_{\nu_I}^0-E_{\nu_I}+i\gamma_{\nu_I}}
\nonumber\\
&= {1\over2}e^{-\kappa_I \Delta T+{\sigma_{t, \tx{S}+\tx{D}}\over2}\kappa_I^2}
\operatorname{erfc}\sqbr{\kappa_I \sigma_{t, \tx{S}+\tx{D}}-\Delta T \over \sqrt{2\sigma_{t, \tx{S}+\tx{D}}}},
\end{align}
where
\al{
\gamma_{\nu_I} &:= {m_{\nu_I}\Gamma_{\nu_I}\over2E_{\nu_I}\fn{\bs{p}_{\nu_I}}},&
\kappa_I &:= \gamma_{\nu_I} + i \pn{E_{\nu_I}\fn{\bs{p}_{\nu_I}} - P^0_{\nu}\fn{\bs{p}_{\nu_I}}},
}
and we used
\al{
{i\over p_{\nu_I}^0-E_{\nu_I}+i\gamma_{\nu_I}}=\int_0^\infty ds\, e^{i s \pn{p_{\nu_I}^0-E_{\nu_I}+i\gamma_{\nu_I}}}.
}

The real part of the argument of the $\operatorname{erfc}$ in Eq.~\eqref{eq: erfc smoothing} changes its sign at $\Delta T = \gamma_{\nu_I} \sigma_{t, \tx{S}+\tx{D}} \simeq \Delta T_{\mr{th},I}\fn{\bs{p}_{\nu_I}}$ given in Eq.~\eqref{eq: time threshold approx}.
The two asymptotic forms, $\operatorname{erfc}\fn{x} \to e^{-x^2}/\sqrt{\pi} x$ for $x \to +\infty$ and $\operatorname{erfc}\fn{x} \to 2$ for $x \to -\infty$, reproduce the saddle-point (virtual) and pole (real) contributions at early and late times, respectively.
In the early-time limit, $\Delta T_{\mr{th},I}\fn{\bs{P}_{\nu}} - \Delta T \gg \sqrt{\sigma_{t, \tx{S}+\tx{D}}}$, the function~\eqref{eq: erfc smoothing} approaches the value at the saddle point~\eqref{eq:pzero-ast},
\al{
\label{eq: saddle asymptotic}
{i\over\sqrt{2\pi\sigma_{t, \tx{S}+\tx{D}}}}\,
{1
\over
p^0_{\nu_I \star}-E_{\nu_I}+i\gamma_{\nu_I}}
e^{-{\pn{\Delta T}^2 \over 2\sigma_{t, \tx{S}+\tx{D}}}},
}
while in the late-time limit, $\Delta T - \Delta T_{\mr{th},I}\fn{\bs{P}_{\nu}} \gg \sqrt{\sigma_{t, \tx{S}+\tx{D}}}$, it approaches the residue of the positive-energy pole,
\al{
e^{-{\sigma_{t, \tx{S}+\tx{D}}\over2} \pn{E_{\nu_I}-i\gamma_{\nu_I}-P^0_{\nu}}^2}\,
e^{-i \pn{E_{\nu_I}-i\gamma_{\nu_I}-P^0_{\nu}} \Delta T}.
}
The factor $e^{-\gamma_{\nu_I}\Delta T}$ in the second exponential represents the decay of the propagating particle and accounts for the exponential suppression of the curves with nonzero $\gamma_{\nu_I}$ at large $\Delta T$ in Fig.~\ref{fig: erfc crossover}.
For the configuration $E_{\nu_I} = P^0_{\nu}$ shown in Fig.~\ref{fig: erfc crossover}, the denominator in Eq.~\eqref{eq: saddle asymptotic} becomes $p^0_{\nu_I \star} - E_{\nu_I} + i\gamma_{\nu_I} = i \pn{\gamma_{\nu_I} - \Delta T/\sigma_{t, \tx{S}+\tx{D}}}$.
Consequently, the function~\eqref{eq: erfc smoothing} and both the asymptotic forms are real, and the saddle-point value~\eqref{eq: saddle asymptotic} diverges at $\Delta T = \Delta T_{\mr{th},I}$.
This divergence is nothing but the $t$-channel singularity in the saddle-point contribution discussed in Sec.~\ref{sec: Real and Virtual Propagation of intermediate particles}, and the exact expression~\eqref{eq: erfc smoothing} renders it finite.
The step function in Eq.~\eqref{eq: flavor changing amplitude} is thereby smoothed into a crossover centered at $\Delta T \simeq \Delta T_{\mr{th},I}\fn{\bs{P}_{\nu}}$ with width $\sqrt{\sigma_{t, \tx{S}+\tx{D}}}$, where $\bs{p}_{\nu_I}$ is evaluated at the saddle point~\eqref{eq:P-spatial_nu}.

At the threshold, the argument of the $\operatorname{erfc}$ in Eq.~\eqref{eq: erfc smoothing} vanishes for the configuration in Fig.~\ref{fig: erfc crossover}, and $\operatorname{erfc}\fn{0} = 1$, so that $J_I\fn{\Delta T_{\mr{th},I}} = \tfrac{1}{2} e^{-n^2/2}$, where $n := \gamma_{\nu_I} \sqrt{\sigma_{t, \tx{S}+\tx{D}}} \simeq \Delta T_{\mr{th},I}/\sqrt{\sigma_{t, \tx{S}+\tx{D}}}$, giving $1/2$, $0.16$, and $0.0056$ for the three curves in Fig.~\ref{fig: erfc crossover}. The separation of the threshold from the origin, $n$, thus also fixes the magnitude of the amplitude there.
Note that $n$ can take any value while the narrow-width condition $m_{\nu_I} \Gamma_{\nu_I} \ll E_{\nu_I}^2$ for Eq.~\eqref{eq: time threshold approx} is maintained.

The effect of the threshold is observable as the deviation of Eq.~\eqref{eq: erfc smoothing} from the two asymptotic forms.
The three curves illustrate the three resulting regimes.
For $n \ll 1$ (leftmost curve), the crossover band is at $\Delta T = 0$, and pure real propagation is established for $\Delta T \gtrsim \sqrt{\sigma_{t, \tx{S}+\tx{D}}}$.
For $n \simeq 1$ (middle curve), the crossover occurs at $\Delta T \sim \Delta T_{\mr{th},I} \sim \sqrt{\sigma_{t, \tx{S}+\tx{D}}}$ with an $O\fn{1}$ amplitude, and the threshold leaves its largest imprint. This is the regime anticipated for sufficiently broad resonances and BSM states in Sec.~\ref{sec: Neutrino Oscillation in WPQFT}.
For $n > 1$ (rightmost curve), the threshold is well separated from the origin, but the amplitude there is already exponentially suppressed by $e^{-n^2/2}$. The asymptotic form~\eqref{eq: Jacob-Sachs theorem} thus describes only this exponentially small tail of the amplitude.

For SM neutrinos, we have $n \lesssim 10^{-59}$, and the corresponding curves are indistinguishable from the leftmost stable-limit curve in Fig.~\ref{fig: erfc crossover}.
In particular, the stepwise structure in $\Delta T_0$ in Eq.~\eqref{eq: stepwise probability} can be resolved only if the separations between the thresholds $\Delta T_{\mr{th},I}\fn{\bs{P}_{\nu}}$ exceed the crossover width $\sqrt{\sigma_{t, \tx{S}+\tx{D}}}$ in Eq.~\eqref{eq: erfc smoothing}, and no such interval exists for the three SM neutrinos.

\subsection{Comments on typical size of wave packet}

In formalisms utilizing wave packets, the sizes of the wave packets for external particles are parameters of the theory.
We will outline several basic points regarding this matter:
(i) As mentioned in Eq.~\eqref{eq:averaged-signma}, for each of the initial and final states, the smallest wave packet size makes the largest contribution to the average wave packet size ($\sqrt{\sigma_\SD}$);
(ii) If a state involving external lines is a bound state, the wave packet size may be regarded as being roughly equal to the average radius of its profile;
(iii) If a state involving external lines is a scattering state, accurately estimating the size of the wave packet is not easy;
(iv) An examination of the mean free path of a certain process may place an upper bound on the wave packet size of the associated external particles;
(v) In our current calculations, we assume that the wave packet size of the external particles is relatively large---that is, not extremely far removed from the plane wave---and incorporate only the leading terms into the computation. The validity of this approximation requires that the wave packet size satisfy a certain lower bound. For instance, in the case of non-relativistic external particles, the de Broglie wavelength serves as a good estimate for this lower bound (see~\cite{Ishikawa:2023bnx} for concrete discussions).
As a supplementary note, it is possible to systematically incorporate terms beyond the leading-order contribution to the wave packet size; indeed, the inclusion of these higher-order terms is essential for analyzing the attenuation effects associated with wave packet propagation. However, regarding this specific point---which remains a subject for future research---we will not engage in a detailed discussion within the scope of this paper;
(vi) A systematic estimation of the size of a scattering wave packet, including its associated uncertainties, is not an easy task. This constitutes an intriguing research topic.

Furthermore, we briefly review recent experimental analyses and theoretical studies concerning the wave packet size of neutrinos.
One approach is to place constraints on the quantity of the effective spatial size (with a length dimension) $\varsigma_x := \sqrt{2\sigma_{\tx{S}+\tx{D}}}$ by comparing the effects of the decoherence term in neutrino oscillations with experimental data (see Eq.~\eqref{eq: Standard Formula}).\footnote{
In the papers referenced below, $\varsigma_x$ is denoted as $\sigma_x$ or $\sigma$; to avoid confusion with our own notation, we have chosen a different font.
}
The Daya Bay experiment imposed $\varsigma_x \gtrsim 10^{-4}\,\tx{nm}$ ($95\%\,\tx{C.L.}$)~\cite{DayaBay:2016ouy},
while a combined analysis by theorists of the data of the Daya Bay~\cite{Cao:2016vwh, DayaBay:2022eyy}, RENO~\cite{RENO:2010vlj, RENO:2012mkc} and KamLAND~\cite{KamLAND:2010fvi} experiments yields $\varsigma_x > 2.1 \times 10^{-4}\,\tx{nm}$ ($90\%\,\tx{C.L.}$)~\cite{deGouvea:2021uvg}.
Note that a future target of the (now ongoing) JUNO experiment was estimated in~\cite{deGouvea:2020hfl}, where it would be sensitive to $\varsigma_x < 2.1 \times 10^{-3}\,\tx{nm}$ ($90\%\,\tx{C.L.}$).
Here, we wish to reiterate that, within the interpretation of our formalism, we are collectively measuring the effects of the initial and final-state wave packets through the decoherence of neutrino oscillations.

Another approach is to estimate the size of the neutrino in beta decays directly by focusing on precise measurements of the low-energy recoiling atoms in nuclear electron capture decay.
The BeEST experiment has investigated the process $e^- + {}^7\tx{Be} \to {}^7\tx{Li} + \nu_e$, which is the simplest pure electron-capture decaying process~\cite{Leach:2021bvh}.
In~\cite{Smolsky:2024uby}, the first direct limit on the spatial width of an external neutrino wave packet is imposed from the energy width of the recoiling nucleus in the decay.
The extracted lower limit on the spatial width of the nuclear recoil is $\varsigma_{\tx{N}, x} \geq 6.2\,\tx{pm}$.\footnote{
$\varsigma_{\tx{N}, x}$ is calculated as follows.
First, the ${}^7\tx{Li}$ recoil spectrum of the $1s$-${}^7\tx{Be}$ to the ground state of ${}^7\tx{Li}$ transition was experimentally measured and its peak energy $E$ and the error $\varsigma_E$ were obtained.
The formula $\varsigma_p = \sqrt{m/(2E)} \, \varsigma_E$ was used to estimate an upper limit on $\varsigma_{\tx{Li}, p}$, and the Heisenberg uncertainty relation gave us the corresponding lower limit on $\varsigma_{\tx{N},x}$.
}
To pull out a limit on the neutrino wave packet size $\varsigma_{\nu,x}$, two approaches were adopted, where
one is based on a conservation of energy~\cite{Akhmedov:2022bjs} and the other is based on a conservation of momentum~\cite{Jones:2022hme}.
The corresponding results are $\varsigma_{\nu, x} \geq 35\,\tx{nm}$ ($95\%\,\tx{C.L.}$) and $\varsigma_{\nu, x} \geq 6.2\,\tx{pm}$ ($95\%\,\tx{C.L.}$), respectively, revealing an enormous spread between the two estimates.
Regarding such a kind of discrepancy, it appears to have been the subject of considerable discussion in recent years~\cite{Beuthe:2001rc, Akhmedov:2022bjs, Jones:2022cvh, Akhmedov:2022mal, Jones:2022hme, Krueger:2023skk, Jones:2024jnp}; however, there still does not seem to be a consensus.
At this juncture, we offer our perspective.
When treating external-line wave packets within our formulation, neither the law of energy conservation nor that of momentum holds strictly true at the point of interaction~\cite{Ishikawa:2018koj}.
This is a natural consequence when considering the inherent uncertainties in position and momentum associated with wave packets. Consequently, estimating the size of the neutrino wave packet involved in the aforementioned electron-capture beta decay using our formalism appears to be an intriguing problem (see also Ref.~\cite{Ishikawa:2025uwl}).

\section{Summary and Discussions}
\label{sec: Discussions and Summary}
In this work, we revisited flavor oscillation phenomena within the framework of wave-packet quantum field theory (WPQFT). By explicitly evaluating the propagation amplitude, we clarified how the asymptotic real-particle behavior described by the Jacob--Sachs theorem~\cite{Jacob:1961zz} emerges
at sufficiently long propagation times. We also showed that this behavior is absent at sufficiently short propagation times, where the intermediate state does not yet behave as a real propagating particle.

This transition has a simple interpretation in the saddle-point analysis. The relevant criterion is determined by comparing the imaginary part of the saddle point of the internal energy, which grows in proportion to the propagation time, with the imaginary part of the pole of the propagator. Once the former exceeds the latter, the deformation of the integration contour crosses the pole, and the real-propagation contribution appears. Conversely, for sufficiently short propagation times, the contour does not cross the pole. In this regime, the amplitude is described only by the saddle-point contribution, which corresponds to virtual propagation.

We found that the transition between the two regimes is a smooth crossover whose center is the threshold $\Delta T_{\mr{th},I} \simeq \gamma_{\nu_I} \sigma_{t, \tx{S}+\tx{D}}$ and whose width is $\sqrt{\sigma_{t, \tx{S}+\tx{D}}}$, the inverse of the energy uncertainty (Sec.~\ref{subsec: Smoothing of the propagating time threshold}). Since flavor transitions during virtual propagation are suppressed for relativistic particles, the standard flavor-oscillation formula describes the behavior after this crossover. For SM neutrinos, the center lies deep inside the width, making the crossover indistinguishable from its stable limit, and the onset time scale is set by $\sqrt{\sigma_{t, \tx{S}+\tx{D}}}$ alone. An observable deviation from the two asymptotic forms requires $\gamma_{\nu_I} \sqrt{\sigma_{t, \tx{S}+\tx{D}}} \simeq 1$, as may occur for unstable intermediate particles or BSM states. 

Conversely, our result can be interpreted as clarifying the range of applicability of the Jacob--Sachs theorem~\cite{Jacob:1961zz}, which is known as an asymptotic formula, and as showing that its validity is already
established
even at extremely short propagation distances. In addition, it is worth mentioning that, compared with previous studies, our analysis is based on a calculation using the general configuration of the central momenta and positions of wave packets.

Although we have kept (sterile) neutrino oscillations in mind as a representative example, the same calculation can be applied straightforwardly to flavor oscillations of mesons and 
BSM particles. In addition, while this work considered only a single internal neutrino line, the generalization to processes involving two or more propagating neutrinos, such as $Z \to \nu \bar{\nu}$, is not difficult. In our analysis, we have neglected the effect of the time-dependent spreading of each wave packet. A detailed analysis of this effect is left for future work.

Before closing, let us raise points concerning the integration over the propagation time of the wave packet. Following previous studies~\cite{Kiers:1995zj, Giunti:1997sk, Giunti:1997wq, Beuthe:2001rc, Akhmedov:2009rb}, we introduced an integration over the propagation time. The rationale was that experiments do not directly measure the propagation time. However, 
this procedure appears to be an a posteriori prescription, and the theoretical origin of this integration has not been well understood. Moreover, it is also unclear whether this integration remains necessary in situations where the propagation time can be measured with high precision.

Furthermore, in performing the propagation-time integration, we regard the localization factor $\psi_X$ at the interaction point as unity. This corresponds to choosing a particular configuration in which, for the central positions of the external wave packets, the saddle-point trajectories of the wave packets intersect at a single point precisely at the time corresponding to the saddle point of the interaction time. If this assumption is relaxed, how is the neutrino-oscillation formula modified?

As we will show in our subsequent work~\cite{Nishiwaki:toappear}, these two questions can be clarified by carefully evaluating the probability formula in WPQFT.

\begin{acknowledgments}
We are grateful to Joachim Kopp, Apostolos Pilaftsis, and Thomas Schwetz for valuable comments on the first arXiv version of this manuscript.
K.N.\ thanks Kazunori Kohri for asking an interesting question about the size of neutrino wave packets.
Furthermore, K.N.\ expresses his gratitude to the organizers of the Majorana--Raychaudhuri Seminar Series for providing him with the opportunity to present an online talk on this research at a stage just prior to its completion.
This work was supported in part by JSPS KAKENHI Grant Numbers 25KJ0022 (JW) and 26K00623 (KO), and by MEXT SPReAD Program Grant No.\,26279286 (KO). The diagrams in this paper were drawn with TikZ-Feynman~\cite{Ellis:2016jkw}.
\end{acknowledgments}

\appendix
\begin{widetext}
\section*{Appendix}
\section{Brief summary on our notation and formalism} \label{app:brief-summary}

Here, we provide a brief summary of our notation and concept to describe wave-packet to wave-packet transitions in our Gaussian wave-packet formalism.
If readers are interested in further details, see~\cite{Ishikawa:2018koj, Ishikawa:2020hph}.
\begin{itemize}
\item
In our formalism, an external state $\psi$ is defined as taking the Gaussian profile at a specific time $X^0_\psi$, with a center coordinate of $\bs {X}_\psi$, a most probable momentum of $\bs{P}_\psi$, and a width of $\sqrt{\sigma_\psi}$.
Furthermore, information on spin is added as needed.
This state constitutes a solution to the corresponding free equations of motion, and its time evolution can be described by the corresponding free Hamiltonian.
See App.~\ref{app: Single-field Gaussian state and Contraction} for our notation for Gaussian quantum states.
\item
Its most distinctive feature is the introduction, in a manifest form, of the degrees of freedom associated with the position coordinates of external particles. Consequently, the measure of the final-state phase space takes the form $d^6 \bs{\Pi} = d^3 \bs{X} d^3 \bs{P}/(2\pi)^3$.
Note that the methodologies adopted, e.g., in references~\cite{Beuthe:2001rc, Akhmedov:2010ms}, do not incorporate information regarding where (at what time(s)) normalizable external states are localized in manifest ways.
\item
The external profile within the (Gaussian) S-matrix is obtained as shown in Eq.~\eqref{eq:psihat-Nstate} (if only the leading part of the $\sigma$-expansion is retained, see App.~\ref{app: Single-field Gaussian state and Contraction} for details).
We may also expand a field by the creation and annihilation operators of the above-mentioned free Gaussian wave packets (as an expansion based on $d^6 \bs{\Pi}$).
Also, the Gaussian basis follows the completeness relation (as an expansion based on $d^6 \bs{\Pi}$), and thus, we can represent intermediate states in the language of the plane wave. Refer to~\cite{Ishikawa:2018koj, Ishikawa:2020hph} for further technical details.
\item
The energy and momentum of an external free Gaussian state are expressed in the usual form, as shown in Eqs.~\eqref{eq:def_energy} and \eqref{eq:def_velocity}, where the corresponding momentum $\bs{p}_\tx{ext}$ (for the state labeled by $\psi$) is integrated.
Here, under the approximation of retaining only the leading term of the $\sigma$-expansion, this $\bs{p}_\tx{ext}$-integral can be evaluated using the saddle-point method.
Here, the saddle-point value of the coordinate momentum retains $\bs{P}_\psi$ as its real part, but acquires a correction in its imaginary part (see Eq.~\eqref{appeq: saddle point of p}).
The symbol appearing here $\bs{X}_{\medstar \psi}\fn{x^0}$ represents the classical trajectory of the center of
the corresponding wave packet at the reference time $x^0$.
\item
In addition to the momentum integrals involved in connecting each external line to the S-matrix, in the case where there is a single intermediate state---as discussed in this paper---the S-matrix includes integrals over the two interaction points, $x^\mu$ and $y^\mu$, as well as over the four-momentum $p^\mu$ associated with the plane-wave expansion of the intermediate state.
Here, Eq.~\eqref{eq: effective propagator} represents the result of integrating the two interaction points using the saddle-point method.
\item 
Equation~\eqref{eq: effective propagator} corresponds to the S-matrix of the overall process after integrating over the interaction points.
Here, as is evident from the absence of delta functions, in our wave-packet formalism, both energy and momentum are conserved at each interaction point only approximately.
This is plausible, considering the properties of wave packets.
See~\cite{Ishikawa:2018koj, Ishikawa:2020hph} for more details.
\item 
The quantities $\bs{L}$ and $\Delta T$ denote the vector representing the propagation distance and propagation time of the intermediate particle, defined in Eq.~\eqref{eq:Delta-variables} as
$\bs{L} = \bs{X}_{\tx{D}} - \bs{X}_{\tx{S}}$,
$L = |\bs{L}|$,
$\Delta T = T_{\tx{D}} - T_{\tx{S}}$.
A major difference from the standard treatment in QFT is that the values of $\bs{X}_{\tx{D}}$, $\bs{X}_{\tx{S}}$, $T_{\tx{D}}$, and $T_{\tx{S}}$ are {\it not} independent degrees of freedom but rather functions of the external states' configuration.
Here, it is highly convenient to use the notation defined in Eq.~\eqref{overline notation} for `$\sigma$-weighted' variables for the source and detector parts, separately.
This implies that by appropriately integrating the squared absolute value of the S-matrix over the canonical phase space of the external particles $\Pi_{w} d^6\bs{\Pi}_w$ ($w$: states), one can calculate the effective neutrino oscillation propagation distance and transition time as functions of unintegrated variables of the external states.
However, this paper will not delve deeply into this specific aspect.
Note that, in other QFT calculations, for instance, in~\cite{Beuthe:2001rc, Akhmedov:2010ms}, $\bs{X}_{\tx{D}}$, $\bs{X}_{\tx{S}}$, $T_{\tx{D}}$, and $T_{\tx{S}}$ are introduced as independent degrees of freedom.
\item 
We have formulated this by generalizing the number of initial and final states.
In this context, the notation that assigns positive and negative signs to incoming and outgoing states, when introduced in conjunction with Eq.~\eqref{eq:pmC-operation}, proves highly effective in facilitating a systematic description.
Note that the symbol defined in Eq.~\eqref{appeq: pm SD notation} is useful for compensating for the difference in some signs between the source and the detector domains.
Note that several related studies formulate the shape of the wave packet in a general form, yet treat the overall reaction as a minimal two-body to two-body process.
\item
$\sqrt{\sigma_{\tx{S}}}$ and $\sqrt{\sigma_{\tx{D}}}$ (defined in Eq.~\eqref{eq:averaged-signma}) describe the average spatial extent of the source and detector regions, whereas $\sqrt{\sigma_{t, \tx{S}}}$ and $\sqrt{\sigma_{t, \tx{D}}}$ (defined in Eq.~\eqref{eq:def_sigma_tSD}) describe their average extent in the temporal direction.
These inverses provide estimates of the average errors in the corresponding momentum and energy, respectively.
\item 
The secondary symbols, $\sigma_{\tx{S}+\tx{D}} = \sigma_{\tx{S}} + \sigma_{\tx{D}}$ (in Eq.~\eqref{eq:sigma_SplusD}) and $\sigma_{t, \tx{S}+ \tx{D}} = \sigma_{t, \tx{S}} + \sigma_{t, \tx{D}}$ (in Eq.~\eqref{eq:sigma_tSplusD}) describe the squares of the average spatial and temporal sizes of the two interactions, respectively.
Here, $\sqrt{2 \sigma_{\tx{S}+\tx{D}}}$ provides a length scale characterizing the decoherence effect of neutrino oscillation (which is proportional to $L$), refer to the standard formula in Eq.~\eqref{eq: Standard Formula}.
Furthermore, as shown in Eq.~\eqref{eq: time threshold}, $\sqrt{\sigma_{t, \tx{S}+ \tx{D}}}$ plays a significant role in determining the timescale for on-shell amplitudes to emerge.
\item 
In estimating the four-dimensional momentum integral for intermediate states, we adopt a strategy in which we first evaluate the energy component, incorporating pole contributions, using the saddle-point method, and subsequently evaluate the momentum component using the same method.
The respective results are summarized in Eqs.~\eqref{eq:pzero-ast} and \eqref{eq:P-spatial_nu}, respectively.
Note that this most probable energy value incorporates the degrees of freedom associated with $\bs{p}_{\nu_I}$ (which are finally integrated out); however, if one wishes merely to observe the primary effects, it suffices to simply replace this with the most probable momentum value, $\bs{P}_{\nu}$.
\item 
The most probable value for the energy of the intermediate state (as discussed in the preceding item) incorporates the degrees of freedom associated with $\Delta T$.
Although $\Delta T$ should, in principle, be uniquely determined by the configuration of the external states, if the phenomenological assumption underlying this paper (namely, that the propagation distance of the intermediate state is sufficiently long) holds, then the value estimated from classical propagation, given in Eq.~\eqref{eq: classical propagation}, should provide a sufficiently good approximation of it.
\end{itemize}

\section{Single-field Gaussian state and Contraction}
\label{app: Single-field Gaussian state and Contraction}
In this Appendix, we summarize the rules for the contraction between fields and external-line states represented by wave packets in the Gaussian WPQFT formalism.
Note that we adopt the mostly plus metric $\paren{-,+,+,+}$ and the other conventions of~\cite{Oda:2023qek}.

As a representative example, we consider the following fermion field as the field to be contracted:
\begin{align}
\widehat\psi \fn{x}
	=	\sum_{s} 
    &\int{d^3 \bs p\over \sqrt{2E_{\psi}\fn{\bs p}}}
    \Bigg[
			{e^{i p_{\psi} x} \over \pn{2\pi}^{3\over2}}\, u_{\psi} \fn{\bs p,s}\widehat a_\psi \fn{\bs p,s}
        +{e^{- i p_{\psi} x} \over \pn{2\pi}^{3\over2}}\,v_\psi\fn{\bs p,s}  \widehat b_{\psi}^\dagger\fn{\bs p,s}
			\Bigg],
\end{align}
where $u_{\psi}\fn{\bs p,s}$ and $v_{\psi}\fn{\bs p,s}$ are plane-wave solutions of the Dirac equation, and $a_{\psi}~(a_{\psi}^{\dagger})$ and $b_{\psi}~(b_{\psi}^{\dagger})$ are the annihilation and creation operators of the plane-wave particle and antiparticle, respectively.
$p_\psi^\mu$ denotes $(E_\psi = \sqrt{m_\psi^2 + \bs p^2}, \bs p)^\mu$.\footnote{
We adopt the following convention for Dirac fields,
\begin{align}
(i\slashed{p}+m_{\psi})u_{\psi}(p,s)&=0,&
(i\slashed{p}-m_{\psi})v_{\psi}(p,s)&=0,& \notag \\
\left\{ \widehat a_\psi\fn{\bs p,s}, \widehat a_{\psi'}^\dagger\fn{\bs p',s'} \right\}
	&=	\delta_{\psi \psi'}\delta_{s s'}\delta^3\fn{\bs p-\bs p'},&
\left\{ \widehat b_\psi \fn{\bs p,s}, \widehat b_{\psi'}^\dagger\fn{\bs p',s'} \right\}
	&=	\delta_{\psi \psi'}\delta_{ss'}\delta^3\fn{\bs p-\bs p'},&
\text{others}
	&=	0.& \notag
\end{align}
}

Next, we introduce the state,
\begin{align}
\Ket{\mc N_{\SD+}}
    &:=\ket{X_{\mc N_{\SD+}}, \bs P_{\mc N_{\SD+}}, S_{\mc N_{\SD+}}; \sigma_{\mc N_{\SD+}}}, \\
\Bra{\mc N_{\SD-}}
    &:=\bra{X_{\mc N_{\SD-}}, \bs P_{\mc N_{\SD-}}, S_{\mc N_{\SD-}}; \sigma_{\mc N_{\SD-}}},
\end{align}
which represents a particle excitation state described by a wave packet corresponding to the field $\psi$ introduced in the main text. We recall that the subscript $\SD$ indicates that the particle belongs to the vertex in the source (detection) region, while the subscript $\pm$ distinguishes the incoming and outgoing particles.
$X_{\mc N_{\SD\pm}}$ stands for $X^\mu_{\mc N_{\SD\pm}} = \left( X^0_{\mc N_{\SD\pm}}, X^i_{\mc N_{\SD\pm}} \right)$, and
$X^i_{\mc N_{\SD \pm}}$, $P^i_{\mc N_{\SD \pm}}$, and $\sigma_{\mc N_{\SD \pm}}$ denote the central position, central momentum, and width of the wave packet associated with each particle at the corresponding time $X^0_{\mc N_{\SD \pm}}$, respectively.

In addition, in this appendix only, we introduce the antiparticle excitation state 
\begin{align}
\Ket{\mc N^{c}_{\SD+}}
    &:=\ket{X_{\mc N^c_{\SD+}}, \bs P_{\mc N^c_{\SD+}}, S_{\mc N^c_{\SD+}}; \sigma_{\mc N^c_{\SD+}}}, \\
\Bra{\mc N^{c}_{\SD-}}
    &:=\bra{X_{\mc N^c_{\SD-}}, \bs P_{\mc N^c_{\SD-}}, S_{\mc N^c_{\SD-}}; \sigma_{\mc N^c_{\SD-}}},
\end{align}
to distinguish particles from antiparticles. Here, we adopt a slightly sloppy notation: on the right-hand side, we do not explicitly write labels indicating antiparticles. The distinction between particles and antiparticles can instead be identified from the subscripts ${\mc N^{(c)}_{\SD\pm}}$ attached to the parameters characterizing the wave packet and to the spin labels.

The expressions of wave-packet one-particle states are given by their representation in momentum space. We assume a Gaussian form, which is often adopted in the literature,
(where, in the following mathematical expression for the Kronecker delta, the distinction between `$+$' and `$-$' is to be disregarded.):
\begin{align}
&\Braket{\bs{p}_{\mc N^{(c)}_{\SD-}}, s^{(c)}_{\mc N_{\SD-}}|\mc N^{(c)}_{\SD+}}   
		:=	\delta_{
            s^{(c)}_{\mc N_{\SD-}}
            S_{\mc N^{(c)}_{\SD+}}
                }
             N_{\mc N^{(c)}_{\SD+}}\exp\!\left[G_{\mc{N}^{(c)}_{\SD+}}\fn{0, \bs{p}_{\mc N^{(c)}_{\SD-}}}\right],\\
&\Braket{\mc N^{(c)}_{\SD-} |\bs{p}_{\mc N^{(c)}_{\SD+}}, s^{(c)}_{\mc N_{\SD+}}} 
		:=	\delta_{
            s^{(c)}_{\mc N_{\SD+}}
            S_{\mc N^{(c)}_{\SD-}}}
             N_{\mc N^{(c)}_{\SD-}}\exp\!\left[G_{\mc{N}^{(c)}_{\SD-}}\fn{0, \bs{p}_{\mc N^{(c)}_{\SD+}}}\right],
\end{align}
where $G_{\mc N^{(c)}_{\SD \pm}}\fn{x, \bs p}$ is Gaussian factor given in Eq.~\eqref{eq: Gaussian exponent} and 
\begin{align}
N_{\mc N^{(c)}_{\SD\pm}} := \pn{\sigma_{\mc N^{(c)}_{\SD\pm}} \over\pi}^{3\over4},
\end{align}
is a normalization factor.
As for the exponential factor, as mentioned in the main text, the expression is identical for both the particle and antiparticle cases.

The momentum eigenstates of the particle and antiparticle are defined as follows:
\begin{align}
\label{appeq: spinor plane waves}
&\ket{\bs{p}_{\mc N_{\SD+}}, s_{\mc N_{\SD+}}}
	:= \widehat a_{\psi}^{\dagger}
            \fn{\bs{p}_{\mc N_{\SD+}}, s_{\mc N_{\SD+}}}\ket{0}, \\
&\ket{\bs{p}_{\mc N^{c}_{\SD+}}, s_{\mc N^{c}_{\SD+}}}
	:=\widehat b_{\psi}^{\dagger}
            \fn{\bs{p}_{\mc N^{c}_{\SD+}}, s_{\mc N^{c}_{\SD+}}}\ket{0}, \\
&\bra{\bs{p}_{\mc N_{\SD-}}, s_{\mc N_{\SD-}}}
	:=\bra{0} \widehat a_{\psi}
        \fn{\bs{p}_{\mc N_{\SD-}}, s_{\mc N_{\SD-}}}, \\
&\bra{\bs{p}_{\mc N^{c}_{\SD-}}, s_{\mc N^{c}_{\SD-}}}
	:= \bra{0} \widehat b_{\psi}
        \fn{\bs{p}_{\mc N^{c}_{\SD-}}, s_{\mc N^{c}_{\SD-}}}.
\end{align}
Here, again somewhat sloppily, we do not introduce explicit labels that directly distinguish particles from antiparticles, as in the case of the wave-packet states. 

The contractions between the one-particle states represented by wave packets and the field operator $\psi$ or its Dirac adjoint $\overline{\psi}$ (defined as $\overline{\psi} = \psi^\dagger \gamma^0$ with the zeroth form of $\gamma^\mu$) can be summarized as follows:
\begin{align}
\widehat\psi \fn{x} \Ket{\mc N_{\SD+}}
	&=U^{\psi}_{\mc N_{\SD+}}\fn{x},\\
\Bra{\mc N^c_{\SD-}}\widehat\psi \fn{x} 
	&=V^{\psi}_{\mc N^{c}_{\SD-}}\fn{x}, \\
\overline{\widehat\psi} \fn{x} \Ket{\mc N^c_{\SD+}}
	&=\overline{V^{\psi}_{\mc N^c_{\SD+}}}\fn{x},\\
\Bra{\mc N_{\SD-}}\overline{\widehat\psi} \fn{x}
	&=\overline{U^{\psi}_{\mc N_{\SD-}}}\fn{x},
\end{align}
where 
\begin{align}
U^{\psi}_{\mc N_{\SD+}}\fn{x}
        &=	\sum_s \int{d^3 \bs p\over\sqrt{2E_{\psi}\fn{\bs p}}}
		 	{u_\psi \fn{\bs p,s} }
             N_{\mc{N}_{\SD+}}
            \exp\!\left[G_{\mc{N}_{\SD+}}\fn{x, \bs{p}}\right],
\label{appeq: U defined}
\\
V^{\psi}_{\mc N_{\SD-}}\fn{x}
	&=	\sum_s \int{d^3 \bs p\over\sqrt{2E_{\psi}\fn{\bs p}}}
			{v_\psi \fn{\bs p,s}}
             N_{\mc{N}^c_{\SD-}}
            \exp\!\left[G_{\mc{N}^c_{\SD-}}\fn{x, \bs{p}}\right],
\label{appeq: V defined}\end{align}
and 
\begin{align}
\overline{V^{\psi}_{\mc N^c_{\SD+}}}\fn{x}
	&=	\sum_s \int{d^3 \bs p\over\sqrt{2E_{\psi}\fn{\bs p}}}
		 	{\ol{v_\psi }\fn{\bs p,s} }\
             N_{\mc{N}^c_{\SD+}}
            \exp\!\left[G^{\ast}_{\mc{N}^c_{\SD+}}\fn{x, \bs{p}}\right],
\label{appeq: Vbar defined}\\
\overline{U^{\psi}_{\mc N_{\SD-}}}\fn{x}
	&=	\sum_s \int{d^3 \bs p\over\sqrt{2E_{\psi}\fn{\bs p}}}
		{\ol{u_\psi}\fn{\bs p,s} }
             N_{\mc{N}_{\SD-}}
            \exp\!\left[G^{\ast}_{\mc{N}_{\SD-}}\fn{x, \bs{p}}\right].
\label{appeq: Ubar defined}
\end{align}

The contraction formulas~\eqref{appeq: U defined} to \eqref{appeq: Ubar defined} contain Gaussian factors in the integrand, which makes analytic evaluation difficult. Therefore, we evaluate the integrals over the external momenta using the saddle point method. This can be computed immediately by taking derivatives of the exponent of the Gaussian function $G_{\mc N_{\SD \pm}}\fn{x, \bs p}$ assigned to each external line.

The saddle point of the external momentum, $\bs{p}_{\star \mc N^{(c)}_{\SD\pm}}$, is determined up to the accuracy of $\pn{\sigma_{\mc N^{(c)}_{\SD \pm}}}^{-1}$ as
\begin{align}
p_{\star\mc N^{(c)}_{\SD\pm}}^i
= p_{\star\mc N^{(c)}_{\SD\pm}}^i\fn{x^0, \bs{x}}
	& :=	P_{\mc N^{(c)}_{\SD\pm}}^i\pm i{x^i-X_{\medstar\mc N^{(c)}_{\SD\pm}}^{i}\fn{x^0}\over
    \sigma_{\mc N^{(c)}_{\SD\pm}}}
		+ \mc O\fn{\sigma^{-2}}. 
\label{appeq: saddle point of p}
\end{align}
At this level of accuracy, the effect of the wave-packet spreading with time is neglected.

The expressions for the contractions~\eqref{appeq: U defined} to \eqref{appeq: Ubar defined}, evaluated using the saddle point method at this level of accuracy, are given respectively as follows:
\begin{align}
\label{appeq: u approximation}
&U^{\psi}_{\mc N_{\SD+}}\fn{x} 
	\simeq
        {u\fn{\bs p_{\star \mc N_{\SD+}},S_{\mc N_{\SD+}}}}
        \Phi_{\mc N_{\SD+}}\fn{x}, 
			\\
\label{appeq: ubar approximation}
&\overline{U^{\psi}_{\mc N_{\SD-}}}\fn{x}
        \simeq
        {\overline u\fn{\bs p_{\star \mc N_{\SD-}}, S_{\mc N_{\SD-}}}}
        \Phi_{\mc N_{\SD-}}\fn{x}, 
\end{align}
for particles, and
\begin{align}
\label{appeq: v approximation}
&V^{\psi}_{\mc N^c_{\SD-}}\fn{x}
        \simeq
		{v\fn{\bs p_{\star \mc N^c_{\SD-}},S_{\mc N^c_{\SD-}}}}
        \Phi_{\mc N_{\SD-}}\fn{x},\\
\label{appeq: vbar approximation}
&\overline{V^{\psi}_{\mc N^c_{\SD+}}}\fn{x}
        \simeq
		{\overline{v}\fn{\bs p_{\star \mc N^c_{\SD+} +},S_{\mc N^c_{\SD+}}}}
        \Phi_{\mc N_{\SD+}}\fn{x},
\end{align}
for anti-particles. Here, we have introduced
\begin{align}
\Phi_{\mc N^{(c)}_{\tx{R}\pm}}\fn{x} 
&:=
	 {
	 	\pn{\pi \sigma_{\mc{N}^{(c)}_{\tx{R} \pm}}}^{- {3 \over4}} \over 
	 	\sqrt{2 E_{\mc{N}^{(c)}_{\tx{R}\pm}}\fn{\bs{p}_{\star\mc{N}^{(c)}_{\tx{R}\pm}}}} 
			} 
    \exp\!\left[\widetilde{G}_{\mc{N}^{(c)}_{\SD\pm}}\fn{x}\right],
\end{align}
where

\begin{align}
\widetilde{G}_{\mc{N}^{(c)}_{\SD\pm}}\fn{x}
    &=\mp iP^0_{\mc{N}^{(c)}_{\SD \pm}}\pn{x^0-X_{\mc{N}^{(c)}_{\SD \pm}}^0}
	   \pm i\bs P_{\mc{N}^{(c)}_{\SD \pm}} \cdot\pn{\bs x-\bs X_{\mc{N}^{(c)}_{\SD \pm}}} 
    -{1\over2\sigma_{\mc{N}^{(c)}_{\SD \pm}}} \pn{x^i-X_{\medstar \mc{N}^{(c)}_{\SD \pm}}^{i}\fn{x^0}}^2.
\end{align}

From these expressions, we find that, as stated in the main text, the distinction between particles and antiparticles is unnecessary as long as we restrict our attention to the exponential part.

\section{Towards an Effective Propagator}
\label{app: Towards an Effective Propagator}
In this Appendix, we present an explicit derivation of the expression for the effective propagator~\eqref{eq: effective propagator}, which plays a central role in the discussion of particle propagation in WPQFT. To this end, we first write down the expression for the amplitude of a general neutrino propagation process, and then evaluate the integrals over the interaction points and interaction times using the saddle point method.

We consider a general process that produces a single neutrino,
\begin{align}
i_{\tx S+} + j_{\tx S+} + \cdots
    \to \nu_{\alpha} + i'_{\tx S-} + j'_{\tx S-} + \cdots,
\end{align}
and its detection process
\begin{align}
\nu_{\beta} + a_{\tx D+} + b_{\tx D+} + \cdots
    \to a'_{\tx D-} + b'_{\tx D-} + \cdots.
\end{align}
The combined process obtained by merging these two processes is considered.
In WPQFT, this sequence of processes is treated as a single process
that contains one internal neutrino line,
\begin{align}
\label{appeq: interaction at detection region}
i_{\tx S+} + j_{\tx S+} + a_{\tx D+} + b_{\tx D+} + \cdots
	\to	i'_{\tx S-} + j'_{\tx S-} + a'_{\tx D-} + b'_{\tx D-} + \cdots,
\end{align}

The amplitude corresponding to this process takes the following form:
\begin{align}
\mc A_I\fn{\nu_\alpha \to \nu_\beta}
	&=	\Bra{i'_{\tx S-}, j'_{\tx S-}, \cdots, a'_{\tx D-}, b'_{\tx D-}, \cdots}
			T  \left[\pn{i\int d^4x_{\tx{S}}}\pn{i\int d^4x_{\tx{D}}}
				\widehat{\mc L}_\tx{eff}^\tx{S}\fn{x_{\tx{S}}}
				\widehat{\mc L}_\tx{eff}^\tx{D} \fn{x_{\tx{D}}}\right]
			\Ket{i_{\tx S+}, j_{\tx S+}, \cdots, a_{\tx D+}, b_{\tx D+}, \cdots}\nonumber\\
	&\simeq
		-\int d^4x_{\tx{S}}
        \int d^4x_{\tx{D}}\nonumber\\
	&\quad\times
					\sum_I
			\int\frac{d^4p_{\nu_I}}{\pn{2\pi}^4}
					G\fn{p^0_{\nu_I}, \bs{p}_{\nu_I}}
			M_{I, \alpha\beta}\fn{m_{\nu_I},\, p^0_{\nu_I},\, \bs{p}_{\nu_I}}
            \nonumber\\
	&\quad\times
		\left[ 
			e^{- i p_{\nu_I} \cdot x_{\tx{S}}}
			\prod_{\mc{N}_{\tx S\pm} = i_{\tx S +}, j_{\tx S +}, \cdots  i'_{\tx S -}, j'_{\tx S -} \cdots }		\Phi_{\mc N_{\tx S\pm}}\fn{x_{\tx{S}}^0, \bs{x}_{\tx{S}}}
				 \right]	
	\
		\left[ 
			e^{i p_{\nu_I} \cdot x_{\tx{D}}}
		\prod_{\mc{N}_{\tx D\pm} = a_{\tx D +}, b_{\tx D+}, \cdots  a'_{\tx D -}, b'_{\tx D -} \cdots }		\Phi_{\mc M_{\tx D\pm}}\fn{x_{\tx{D}}^0, \bs{x}_{\tx{D}}}
				 \right],
\label{appeq: amplitude after suddle-point integral}
\end{align}
where $T$ denotes the time ordering;
$\widehat{\mc L}_\tx{eff}^\tx{S}\fn{x_{\tx{S}}}$ and $
\widehat{\mc L}_\tx{eff}^\tx{D} \fn{x_{\tx{D}}}$ denote the interaction terms in the source region and detection region; and 
$x_{\SD}^0$ and $\bs{x}_{\SD}$ denote the interaction time and interaction point in the source and detection regions, respectively. In the second line, we have evaluated the contraction between the wave-packet states and the field operators for each particle using the formulas~\eqref{appeq: u approximation}--\eqref{appeq: vbar approximation} derived in the previous section. 

Here, $M_{I, \alpha\beta}\fn{m_{\nu_I},\, p^0_{\nu_I},\, \bs{p}_{\nu_I}}$ denotes the factorized quantity that collects the spinor wave functions and the flavor mixing matrices, as mentioned in the main text. Because the spinor wave functions are evaluated at the saddle point of the external-line momenta, $M_{I, \alpha\beta}\fn{m_{\nu_I},\, p^0_{\nu_I},\, \bs{p}_{\nu_I}}$ does not depend on the interaction points at the leading order in $\Big(\sigma_{\mc N^{{(c)}}_{\SD \pm}}\Big)$, while such dependence appears at higher orders. However, in evaluating the saddle point, we perform the calculation focusing only on the exponential part, except for the internal-line momenta where poles are present.

For example, consider the processes introduced at the beginning of Section~\ref{sec: External Wave Packet Model},
\begin{align}
\tx{n}_{\tx S+} \to \ol{\nu}_e + \tx{p}_{\tx S-} + \tx{e}^{-}_{\tx S-},
\end{align}
and
\begin{align}
\ol{\nu}_e + \tx{p}_{\tx D+} \to \tx{e}^{+}_{\tx D-} + \tx{n}_{\tx D-}.
\end{align}
Taking these as a concrete example, the explicit expression of 
$M_{I, ee}\fn{m_{\nu_I},\, p^0_{\nu_I},\, \bs{p}_{\nu_I}}$ in this case is given as follows:
\begin{align}
&M_{I, ee}\fn{m_{\nu_I},\, p^0_{\nu_I},\, \bs{p}_{\nu_I}} \nonumber\\
	&=	
            U_{\tx eI}U_{\tx eI}^*
			\frac{G_\tx{F}^2 }{2}
			\left[
				\ol{u}_{e}\fn{\bs{p}_{\star\tx{e}S-}\fn{x_{\tx{S}}^0, \bs{x}_{\tx{S}}}, s_{\tx{e}S-}}
				\gamma^\lambda\pn{1-\gamma^5}(-i{\slashed{p}_{\nu_I}} + m_{\nu_I})
				\gamma^\mu\pn{1-\gamma^5}
				v_{e}\fn{\bs{p}_{\star\tx{e}D-}\fn{x_{\tx{D}}^0, \bs{x}_{\tx{D}}}, s_{\tx{e}D-}}
			\right]\nonumber\\
	&\quad\times
			\left[
				\ol{u}_{p}\fn{\bs{p}_{\star\tx{p}S-}\fn{x_{\tx{S}}^0, \bs{x}_{\tx{S}}}, s_{\tx{p}S-}}
				\gamma_\lambda\pn{1-\tilde{g}\gamma^5}
				u_{n}\fn{\bs{p}_{\star \tx{n}S+}\fn{x_{\tx{S}}^0, \bs{x}_{\tx{S}}}, s_{\tx{n}S+}}
				   			\right]\nonumber\\
	&\quad \times
			\left[
				\ol{u}_{n}\fn{\bs{p}_{\star\tx{n}D-}\fn{x_{\tx{D}}^0, \bs{x}_{\tx{D}}}, s_{\tx{n}D-}}
				\gamma_\mu\pn{1-\tilde{g}\gamma^5}
				u_{p}\fn{\bs{p}_{\star\tx{p}D+}\fn{x_{\tx{D}}^0, \bs{x}_{\tx{D}}}, s_{\tx{p}D+}}
			\right],
\end{align}
where $U_{\alpha I}$ denotes P(KMTY)MNS matrix~\cite{Pontecorvo:1957cp, Pontecorvo:1957qd, Katayama:1962mx, Maki:1962mu}, $G_\tx{F}$ is the Fermi constant, 
and $\tilde g\simeq 1.26$.

As with the external momenta, the space-time coordinates associated with the interaction points are difficult to evaluate analytically; therefore, they are evaluated using the saddle-point method. The Gaussian weight that provides localization at the interaction points is given, for the source and detection regions, respectively, by
\begin{align}
\widetilde{G}_{\tx{S}}\fn{x_{\tx{S}}^0, \bs{x}_{\tx{S}}}
	&= \sum_{\mc{N}_{\tx S\pm}} \widetilde{G}_{\mc{N}_{\tx{S}\pm}}\fn{x_{\tx{S}}^0, \bs{x}_{\tx{S}}}, \\
\widetilde{G}_{\tx{D}}\fn{x_{\tx{D}}^0, \bs{x}_{\tx{D}}}
	&= \sum_{\mc{N}_{\tx D\pm}} \widetilde{G}_{\mc{N}_{\tx{D}\pm}}\fn{x_{\tx{D}}^0, \bs{x}_{\tx{D}}},
\end{align}
from which the saddle points are given by the solutions for which the first derivatives of these functions vanish.

More explicitly, the exponent can be expressed as
\begin{align}
\widetilde{G}_{\tx{S}}\fn{x_{\tx{S}}^0, \bs{x}_{\tx{S}}}
	&=
		- ip_{\nu_I}\cdot x_{\tx S}\nonumber\\
	&\quad
		+\sum_{\mc N_{\tx S\pm}}
		\left[
			\mp i P^0_{\mc N_{\tx S\pm}}x_{\tx S}^0 
			\pm i \bs P_{\mc N_{\tx S\pm}}\!\cdot\! \bs x_{\tx S} 
			-{1\over2\sigma_{\mc N_{\tx S\pm}}}
				\pn{x_{\tx S}^i - X_{\medstar \mc N_{\tx S\pm}}^i\fn{x_{\tx S}^0}}^2
		\right]\nonumber\\
	&\quad
		+ \cdots ,\\
\widetilde{G}_{\tx{D}}\fn{x_{\tx{D}}^0, \bs{x}_{\tx{D}}}
	&=
		 ip_{\nu_I}\cdot x_{\tx D}\nonumber\\
	&\quad
		+\sum_{\mc N_{\tx D\pm}}
		\left[
			\mp i P^0_{\mc N_{\tx D\pm}}x_{\tx D}^0 
			\pm i \bs P_{\mc N_{\tx D\pm}}\!\cdot\! \bs x_{\tx D}
			-{1\over2\sigma_{\mc N_{\tx D\pm}}}
				\pn{x_{\tx D}^i - X_{\medstar \mc N_{\tx D\pm}}^i\fn{x_{\tx D}^0}}^2
		\right]\nonumber\\
	&\quad
		+\cdots
\end{align}
where the terms included in $+\cdots$ represent phases that do not depend on the integration variables nor on the neutrino mass, such as 
$- \sum_{\mc{N}_{\tx{S}\pm}}\pn{\pm iP_{\mc{N}_{\tx{S}\pm}}\cdot X_{\mc{N}_\tx{S}\pm}}$ 
and 
$- \sum_{\mc{N}_{\tx{D}\pm}}\pn{\pm i P_{\mc{N}_{\tx{D}\pm}}\cdot X_{\mc{N}_\tx{D}\pm}}$. 
These are unphysical phases, since they cancel out when taking the absolute value of the amplitude.

The first- and second-order derivatives of the exponential factor $\widetilde{G}_{\SD}\fn{x_{\SD}^0, \bs x_{\SD}}$ with respect to the interaction coordinate $x_{\SD}^i$ are
\begin{align}
\label{appeq: first derivative x of tilde G}
{\partial{\widetilde{G}}_{\SD}\fn{x^0_{\SD}, \bs x_{\SD}}\over\partial x_{\SD}^i}
	&=- i\pn{\pn{-1}^{\SD} p_{\nu_I}^i - \ol{P^i}_{\SD 0}} 
		- \sigma_{\SD}^{-1}
			\pn{x_{\SD}^i - \ol{X_{\medstar}^i \fn{x_{\SD}^0}}_{\SD1}} \\
\label{appeq: second derivative x of tilde G}
{\partial^2 \widetilde{G}_{\SD}\fn{x^0_{\SD}, \bs x_{\SD}}\over\partial x_{\SD}^i\partial x_{\SD}^j}
	&=	-\sigma_{\SD}^{-1} \delta^{ij},
\end{align}
respectively. Thus, the saddle point of the position integration over the interaction points associated with these Gaussian weights is given by
\begin{align}
x^i_{\SD\star} 
= x^i_{\SD\star}\fn{x_{\SD}^0}
	= \ol{X_{\medstar}^i \fn{x_{\SD}^0}}_{\SD1} 
		- i \sigma_{\SD} \pn{\pn{-1}^{\SD} p_{\nu_I}^i - \ol{P^i}_{\SD 0}}.
\end{align}

After performing the position integration over the interaction points, the exponent contains a localization factor associated with the interaction time,
\begin{align}
\widetilde{G}_{\SD}\fn{x_{\SD}^0, \bs x_{\SD\star}} 
	&= i\sqbr{ \pn{-1}^{\SD} p_{\nu_I}^0 -  \ol{P^0}_{\SD 0}}  x_{\SD}^0 
				- i\sqbr{ \pn{-1}^{\SD} \bs{p_{\nu_I}} -  
					\ol{\bs{P}}_{\SD 0} }\cdot \ol{\bs{X}_{\medstar} \fn{x_{\SD}^0}}_{\SD1}
			\nonumber\\
	&\quad
			- {1\over2\sigma_{\SD}}
				\sqbr{
					\ol{X^i_{\medstar} \fn{x_{\SD}^0}^2}_{\SD1} 
					- \pn{\ol{X^i_{\medstar} \fn{x_{\SD}^0}}_{\SD1}}^2
					}\nonumber\\
	&\quad
		-{\sigma_{\SD} \over 2} \sqbr{\pn{-1}^{\SD} p_{\nu_I}^i - \ol{P^i}_{\SD 0}}^2 +\cdots,
\end{align}
where $\cdots$ denotes a phase independent of the integration variables.

The first- and second-order derivatives of $\widetilde{G}_{\SD}\fn{x_{\SD}^0, \bs x_{\SD\star}}$ with respect to the interaction time $x^0_{\SD}$ are given by
\begin{align}
\label{appeq: first derivative x0 of tilde G}
{\partial{\widetilde{G}}_{\SD}\fn{x^0_{\SD}, \bs x_{\SD\star}}\over\partial x_{\SD}^0}
    &=  i   \sqbr{\pn{-1}^{\SD} p_{\nu_I}^0 - \ol{P^0}_{\SD 0}} 
			- i \sqbr{\pn{-1}^{\SD}\bs{p_{\nu_I}} - \ol{\bs{P}}_{\SD 0}} \cdot  \ol{\bs{v}}_{\SD 1}  
			-\sigma_{\SD}^{-1} 
				\sqbr{
					 \ol{v^i X^i_{\medstar} \fn{x_{\SD}^0}}_{\SD1}	
					- \ol{v^i}_{\SD1} \ol{X^i_{\medstar} \fn{x_{\SD}^0}}_{\SD1}
				}\nonumber\\
	&=  i   \sqbr{\pn{-1}^{\SD} p_{\nu_I}^0 - \ol{P^0}_{\SD 0} - q_{\SD}\fn{\bs{p}_{\nu_I}}} 
		-\sigma_{t, \SD}^{-1} 
		\pn{
		x^0_{\SD} 
			- T_{\SD}
		},\\
\label{appeq: second derivative x0 of tilde G}
{\partial^2 \widetilde{G}_{\SD}\fn{x^0_{\SD}, \bs x_{\SD\star}}\over \pn{\partial x_{\SD}^0}^2}
	&=	-\sigma_{t, \SD}^{-1}, 
\end{align}
respectively. Therefore, the saddle point with respect to the interaction time is given by
\begin{align}
x^0_{\SD\star} 
	&=T_{\SD}+ i \sigma_{t, \SD}  \pn{\pn{-1}^{\SD} p_{\nu_I}^0 - \ol{P^0}_{\SD 0} - q_{\SD}\fn{\bs{p}_{\nu_I}}}.
\end{align}
Here,
\begin{align}
q_{\SD}\fn{\bs{p}_{\nu_I}}
	&:=  \pn{\pn{-1}^{\SD}\bs{p}_{\nu_I} - \ol{\bs{P}}_{\SD 0}} \cdot  \ol{\bs{v}}_{\SD 1}.
	\label{appeq: def qSD}
\end{align}
We recall $\bs{X}_{\SD} = \ol{\bs{X}_{\medstar}\fn{T_{\SD}}}_{\SD1}$ in Eq.~\eqref{eq: XSD}, also $\bs{L} = \bs{X}_{\tx{D}}-\bs{X}_{\tx{S}}$, $L = |\bs{L}|$, and $\Delta T := T_{\tx{D}} - T_{\tx{S}}$.

From the above results, evaluating the interaction-time integrations using the saddle-point method as well, one finally obtains the following expression for the effective propagator:
\begin{align}
\mc A_{I, \alpha\beta}
	&\simeq
			\int\frac{d^4 p_{\nu_I}}{\pn{2\pi}^4}
					  G\fn{p^0_{\nu_I}, \bs{p}_{\nu_I}}
                      \Psi\fn{p^0_{\nu_I}, \bs{p}_{\nu_I}}  
			e^{
					-  i p_{\nu_I}^0 \Delta T
					+i\bs p_{\nu_I}\cdot\bs L
					+i\varphi
					}
					\nonumber \\
    &\quad \times
			M_{I,\alpha\beta}
            \fn{m_{\nu_I}, \, p^0_{\nu_I},\bs{p}_{\nu_I}},
\label{appeq: effective propagator}
\end{align}
where $G\fn{p^0_{\nu_I}, \bs{p}_{\nu_I}}$ is the propagator given in Eq.~\eqref{eq: neutrino propagator} and
\begin{itemize}
    \item The overlap function $\Psi\fn{p^0_{\nu_I}, \bs{p}_{\nu_I}}$ can be expressed as
    \begin{align}
    \qquad
    \Psi\fn{p^0_{\nu_I}, \bs{p}_{\nu_I}}
        &:=\prod_{\SD=\tx{S}, \tx{D}}
		\pn{2\pi \sigma_{t, \SD}}^{1\over2} \pn{2\pi \sigma_{\SD}}^{3\over2} 
        \left[
        \prod_{\mc{N}_{\tx{R}\pm}}
			{\pn{\pi \sigma_{\mc{N}_{\SD \pm}}}^{- {3 \over 4}}
			\over 
			\sqrt{2 E_{\mc{N}_{\SD}}\fn{\bs{P}_{\mc{N}_{\SD\pm}}}} 
			}
        \right]
            \nonumber\\
	&\phantom{:=\prod_{\SD=\tx{S}, \tx{D}}}
		\times
			 \exp\pn{
			-{\sigma_{\SD} \over 2} \pn{\pn{-1}^{\SD} p_{\nu_I}^i - \ol{P^i}_{\SD0}}^2
				}\nonumber\\
	&\phantom{:=\prod_{\SD=\tx{S}, \tx{D}}} 
		\times 
			\exp\pn{
		- {\sigma_{t, \SD} \over 2}  \pn{\pn{-1}^{\SD} p_{\nu_I}^0 - \ol{P^0}_{\SD 0} - q_{\SD}\fn{\bs{p}_{\nu_I}}}^2
				}\nonumber\\
	&\phantom{:=\prod_{\SD=\tx{S}, \tx{D}}}
		\times
			\exp\pn{
				- {1\over 2\sigma_{\SD}}
				\pn{
					 \ol{X^i_{\medstar} \fn{T_{\SD}}^2}_{\SD1} 
					-  \pn{X^i_{\SD}}^2
					}
				}.
    \label{appeq: internal momentum localization function}
    \end{align}
\item The phase factor $\varphi$, which is independent of the neutrino momentum, is given by
\begin{align}
    \exp\pn{i\varphi } 
        &:= \exp\pn{
				- i    \pn{\ol{P^0}_{\tx{S} 0}  T_{\tx{S}} + \ol{P^0}_{\tx{D} 0}  T_{\tx{D}} }
				+ i \pn{ \ol{\bs{P}}_{\tx{S} 0} \cdot
					 \bs{X}_{\tx{S}}
					+  \ol{\bs{P}}_{\tx{D} 0} \cdot
					 \bs{X}_{\tx{D}}
					}
				- i \pn{ \ol{P \cdot X}_{\tx{S}0} +  \ol{P \cdot X}_{\tx{D}0}}
				}. 
\label{appeq: constant phase}
\end{align}
\end{itemize}

\section{Evaluation of the internal four-momentum integration}
\label{app: Evaluation of the internal-momentum integration}

\subsection{Evaluation of the internal energy integration}
When performing the integration over the internal energy,  we encounter integrals of the following form:
\begin{align}
    \int d p_{\nu_I}^0 \,\exp\!\left[F\fn{p_{\nu_I}^0}\right] 
        G\fn{p^0_{\nu_I}, \bs{p}_{\nu_I}}
        Q\fn{p_{\nu_I}^0},
\end{align}
where
\begin{align}
    F\fn{p_{\nu_I}^0}
        &=-  i p_{\nu_I}^0 \Delta T \nonumber\\
		&\quad -\sumSandD {\sigma_{t, \SD} \over 2}  \pn{\pn{-1}^{\SD} p_{\nu_I}^0 - \ol{P^0}_{\SD 0} - q_{\SD}\fn{\bs{p}_{\nu_I}}}^2,
\end{align}
and $Q\fn{p_{\nu_I}^0}$ denotes the internal-energy--dependent factors other than the propagator with poles, together with the exponential factor that encodes localization and phase; for example, $M_{I,\alpha\beta} \fn{m_{\nu_I}, \, p^0_{\nu_I}, \bs{p}_{\nu_I}}$. In the following calculation, we neglect the effect of the integrand other than the exponent and propagator with poles on the saddle point, since such effects give higher-order contributions.

The first and second derivatives of the exponential factor $F\fn{p_{\nu_I}^0}$ are given by
\begin{align}
{\partial F\fn{p_{\nu_I}^0}\over \partial p_{\nu_I}^0}
	&= 
		 - i \Delta T  
			-  \sigma_{t, \tx{S}+ \tx{D}}  \pn{ 
				p_{\nu_I}^0 
				-P^0_{\nu}
					},
		\label{first derivative of tilde F}\\
{\partial^2 F\fn{p^{0}_{\nu_I}}\over \pn{\partial p^0_{\nu_I}}^2 }
	&=	-\sigma_{t, \tx{S}+ \tx{D}}, 
			\label{appeq: second derivative of tilde F}
\end{align}
where
\begin{align}
\label{appeq: def P0nu}
P^0_{\nu} \fn{\bs{p}_{\nu_I}}
	&:=
          \sumSandD  \pn{-1}^{\SD} { \sigma_{t, \SD} \over  \sigma_{t, \tx{S}+ \tx{D}}}
		 	\ol{P^0}_{\SD 0} \nonumber\\
	&\quad
        + \sumSandD  \pn{-1}^{\SD} { \sigma_{t, \SD} \over  \sigma_{t, \tx{S}+ \tx{D}}}
		 	\pn{\pn{-1}^{\SD}\bs{p}_{\nu_I}  - \ol{\bs{P}}_{\SD 0}} \cdot \ol{\bs{v}}_{\SD 1}.
\end{align}
Then, as presented in the main text, the saddle point of the internal energy is given as follows:
\begin{align}
p_{\nu_I \star}^0
    &= P^{0}_{\nu}\fn{\bs{p}_{\nu_I}} - i \sigma_{t, \mr{S}+\mr{D}}^{-1} \Delta T.
\end{align}

Since we neglect the effect of wave-packet spreading, the second derivative of $F\fn{p_{\nu_I}^0}$ is real and negative. Therefore, the steepest-descent contour is parallel to the real axis. Accordingly, as shown in Fig.~\ref{fig: Integration contour} of the main text, it is necessary to deform the original integration contour along the real axis into a straight line parallel to the real axis that passes through the saddle point. In this deformation, if the contour crosses the poles of $G\fn{p^0_{\nu_I}, \bs{p}_{\nu_I}}$, whose locations are given by
\begin{align}
p_{\nu_I \pm}^0\fn{\bs{p}_{\nu_I}}
    &= \pm \sqrt{\bs{p}_{\nu_I}^2 + m_{\nu_I}^2 - i m_{\nu_I} \Gamma_{\nu_I}},
\end{align}
then the contributions from those poles must be added. The contribution from this pole corresponds to real propagation.

Evaluating the exponent $F\fn{p_{\nu_I}^0}$ at the saddle point, we obtain
\begin{align}
F\fn{p_{\nu_I\star}^0}
    &=-  i P^0_{\nu} \fn{\bs{p}_{\nu_I}} \Delta T \nonumber\\
	&\quad
		- {1 \over 2} \sumSandD
        \sigma_{t, \SD}  
        \pn{\pn{-1}^{\SD} P^0_{\nu} \fn{\bs{p}_{\nu_I}}- \ol{P^0}_{\SD 0} 
        - q_{\SD}\fn{\bs{p}_{\nu_I}}}^2 \nonumber\\
	&\quad	
		- {1\over2\sigma_{t, \tx{S} + \tx{D}}} \Delta T^2.
\end{align}
Therefore, we find that the contribution of the virtual propagation associated with the saddle point contains an exponential factor suppressed by $\Delta T^2$.

\subsection{Evaluation of the internal three momentum integration}
After performing the integration over the internal energy, one must evaluate the momentum integral. In doing so, the contribution from the poles (corresponding to real propagation) and the contribution from the saddle point (corresponding to virtual propagation) should be treated separately. We first consider the real propagation. The following integral must be evaluated:
\begin{align}
\int d^3  \bs{p}_{\nu_I}\,\exp\!\sqbr{H_{\pm}\fn{\bs{p}_{\nu_I}}} R_{\pm} \fn{\bs{p}_{\nu_I}}.
\end{align}
The function $H_{\pm} \fn{\bs{p}_{\nu_I}}$ in the integrand represent localization of internal momentum:
\begin{align}
H_{\pm} \fn{\bs{p}_{\nu_I}}
    &=   
		- i \, p_{\nu_I \pm}^0\fn{\bs{p}_{\nu_I}}
        \Delta T + i p_{\nu_I}^i L^i  \nonumber\\
	&
        \quad
		+\sumSandD \left[-{\sigma_{\SD} \over 2} \pn{\pn{-1}^{\SD} p_{\nu_I}^i - \ol{P^i}_{\SD0}}^2 
			- {\sigma_{t, \SD} \over 2}  
            \pn{\pn{-1}^{\SD}
            p_{\nu_I \pm}^0\fn{\bs{p}_{\nu_I}}
                     - \ol{P^0}_{\SD 0} - q_{\SD}\fn{\bs{p}_{\nu_I}}}^2 \right],
			\label{appeq: H defined}
\end{align}
where the sign $\pm$ indicates the pole contribution that is picked up in the contour deformation: $+$ corresponds to the forward-in-time process ($\Delta T > 0$), in which the pole associated with positive energy is taken, while $-$ corresponds to the backward-in-time process ($\Delta T < 0$), in which the pole associated with negative energy is taken. $R_{\pm} \fn{\bs{p}_{\nu_I}}$ denotes the internal-momentum--dependent factors other than the propagator with poles, together with the exponential factor that encodes localization and phase. Again, we neglect the effect of this factor on the saddle point, since such effects give higher-order contributions.

Then, the first and second derivatives of the exponent~\eqref{appeq: H defined} for forward-in-time process ($\Delta T > 0$) are given by
\begin{align}
{\partial H_{+}\over\partial p_{\nu_I}^i}\fn{\bs{p}_{\nu_I}}
	&\simeq \sumSandD \Bigg[- \sigma_{\SD} \pn{p_{\nu_I}^i -\pn{-1}^{\SD}  \ol{P^i}_{\SD0}} \nonumber\\
		&\phantom{\simeq \sumSandD \Bigg[} 
			-  \sigma_{t, \SD}   \pn{E_{\nu_I}\fn{\bs{p}_{\nu_I}} - \pn{-1}^{\SD} \pn{\ol{P^0}_{\SD 0} + q_{\SD}\fn{\bs{p}_{\nu_I}}}}
			 \pn{ v^i_{\nu_I}\fn{\bs{p}_{\nu_I}} - \ol{v}^i_{\SD 1} } \Bigg] \nonumber\\
	&\quad
	-  i v^i_{\nu_I} \fn{\bs{p}_{\nu_I}}  \Delta T + i  L^i \nonumber\\
	&\simeq- \sigma_{\tx{S}+ \tx{D}} \pn{p_{\nu_I}^i - P^i_{\nu} }
		-  i v^i_{\nu_I} \fn{\bs{p}_{\nu_I}}  \Delta T + i  L^i,
		\label{appeq: first derivative of H+}\\
{\partial^2 H_{+}\fn{\bs p_{\nu_I}}\over \partial p_{\nu_I}^i\partial p_{\nu_I}^j}
	&= -\sigma_{\tx{S}+\tx{D}} \delta^{ij}
		- i{\Delta T\over E_{\nu_I} \fn{\bs p}}\pn{\delta^{ij}-v_{\nu_I}^i\fn{\bs p}v_{\nu_I}^j\fn{\bs p}} + \cdots \nonumber\\
	&\simeq - \sigma_{\tx{S}+\tx{D}} \delta^{ij}.
			\label{appeq: second derivative of H+}
\end{align}
Here, in the first derivative, we have retained only the terms that survive at leading order in $m_{\nu_I}$ and $\Gamma_{\nu_I}$.\footnote{
The term proportional to $\sigma_{\tx{S}+\tx{D}}$ appearing in Eq.~\eqref{appeq: H defined} represents momentum conservation, while the term proportional to $\sigma_{t, \SD}$ reduces, in the plane-wave limit, to delta functions that impose energy conservation and the on-shell condition. In this sense, when evaluating the first and second derivatives, it is sufficient to focus on the term proportional to $\sigma_{\tx{S}+\tx{D}}$, which characterizes momentum localization, and on the imaginary part, as long as the masses can be neglected.} In the second derivative, the imaginary part has been omitted, since it corresponds to the dispersion effect.

Focusing only on the forward-in-time process ($\Delta T > 0$), the saddle point of the momentum integral in the real propagation is given by
\begin{align}
p_{\nu_I \star+}^i
	&:= P_{\nu}^i 
	+ i\,{ L^i -  v^i_{\nu_I} \fn{\bs{P}_{\nu}}\Delta T  \over \sigma_{\tx{S}+ \tx{D}} }
	+ \mathcal{O}(\sigma^{-2}) .
\end{align}
The exponent evaluated at this saddle point, $H_{+} \fn{\bs{p}_{\nu_I \star}}$, is given by
\begin{align}
H_{+} \fn{\bs{p}_{\nu_I \star+}}
	&=
		-  i \, p_{\nu_I +}^0\fn{\bs{P}_{\nu}}  \Delta T + i P_{\nu}^i L^i \nonumber\\
	&\quad
		-{\sigma \over 2}  \pn{\ol{P^i}_{\tx{S}0}+ \ol{P^i}_{\tx{D}0}}^2  \nonumber\\
	&\quad
		- {\sigma_{t} \over 2}  \pn{
							\ol{P^0}_{\tx{S}0} + \ol{P^0}_{\tx{D}0} 
							+ \pn{
								q_{\tx{S}} \fn{\bs{P}_{\nu}} + q_{\tx{D}} \fn{\bs{P}_{\nu}}
								} 
							}^2 \nonumber\\
	&\quad
		- {\sigma_{t, \tx{S}+\tx{D}} \over 2}
					\pn{
						p_{\nu_I +}^0\fn{\bs{P}_{\nu}} - P_{\nu}^0 \fn{\bs{P}_{\nu}}
						}^2 \nonumber\\
	&\quad
		-{1 \over 2\sigma_{\tx{S}+ \tx{D}}}  \pn{ L^i -  v^i_{\nu_I} \fn{\bs{P}_{\nu}}\Delta T}^2.
\end{align}

To derive this expression, we have used the following manipulations:
\begin{align}
\label{appeq: derivation of H0 expression}
-{1 \over 2} \sumSandD \sigma_{\SD} \pn{ P_{\nu}^i- \pn{-1}^{\SD}\ol{P^i}_{\SD0}}^2 
	&=
	- {\sigma_{\tx{S}} \over 2} \pn{P_{\nu}^i - \ol{P}^i_{\tx{S}0}}^2 
		- {\sigma_{\tx{D}} \over 2} \pn{P_{\nu}^i + \ol{P}^i_{\tx{D}0}}^2  \nonumber \\
	&=
		-{\sigma_{\tx S} \over 2} \pn{
					{1 \over \sigma_{\tx{S}+\tx{D}} }
					\pn{
						\sigma_{\tx{D}}  \ol{P}^i_{\tx{S}0}
							+ \sigma_{\tx{S}}  \ol{P}^i_{\tx{D}0}
						}
				}^2
		-{\sigma_{\tx D} \over 2} \pn{
					{1 \over \sigma_{\tx{S}+\tx{D}} }
					\pn{
						\sigma_{\tx{S}}  \ol{P}^i_{\tx{D}0}
							+ \sigma_{\tx{D}}  \ol{P}^i_{\tx{S}0}
						}
				}^2\nonumber \\
	&=
		-{\sigma \over 2}  \pn{\ol{P^i}_{\tx{S}0}+ \ol{P^i}_{\tx{D}0}}^2,
\end{align}
and
\begin{align}
    - {1 \over 2} \sumSandD 
			\left[
			 	\sigma_{t, \SD}  
					\pn{
                    p_{\nu_I +}^0\fn{\bs{P}_{\nu}} 
					- \pn{-1}^{\SD} \pn{\ol{P^0}_{\SD 0} + q_{\SD}\fn{\bs{P}_{\nu}}
					}}^2 
					\right] 
		&= - {1 \over 2} \sumSandD \sigma_{t, \SD}
						\pn{
						P_{\nu}^0\fn{\bs{P}_{\nu}}
						-\pn{
							p_{\nu_I +}^0\fn{\bs{P}_{\nu}} - P_{\nu}^0\fn{\bs{P}_{\nu}}
							}
						-\pn{-1}^{\SD} \pn{\ol{P}_{\SD 0} + q_{\SD} \fn{\bs{P}_{\nu}}}
						}^2\nonumber \\
	&= - {1 \over 2} \sumSandD \sigma_{t, \SD}
			\pn{
				P_{\nu}^0\fn{\bs{P}_{\nu} }
					-\pn{-1}^{\SD} \pn{\ol{P}_{\SD 0} + q_{\SD} \fn{\bs{P}_{\nu}}}
				}^2\nonumber \\
	&\quad
		- {1 \over 2} \sumSandD \sigma_{t, \SD}
			\pn{
				p_{\nu_I +}^0\fn{\bs{P}_{\nu}} - P_{\nu}^0\fn{\bs{P}_{\nu}}
				}^2\nonumber \\
	&\quad
		-\sumSandD \sigma_{t, \SD} 
			\pn{
				P_{\nu}^0\fn{\bs{P}_{\nu} }
					-\pn{-1}^{\SD} \pn{\ol{P}_{\SD 0} + q_{\SD} \fn{\bs{P}_{\nu}}}
				}
			\pn{
				p_{\nu_I +}^0\fn{\bs{P}_{\nu}} - P_{\nu}^0\fn{\bs{P}_{\nu}}
				}\nonumber \\
	&=
		  - {\sigma_{t} \over 2}
		  	 \pn{\ol{P^0}_{\tx{S}0} + \ol{P^0}_{\tx{D}0} 
		  	+\pn{
				q_{\tx{S}}\fn{\bs{P}_{\nu}} + q_{\tx{D}}\fn{\bs{P}_{\nu}} 
				}
				}^2
		   - {\sigma_{t, \tx{S}+\tx{D}} \over 2} 
		   	\pn{
				p_{\nu_I +}^0\fn{\bs{P}_{\nu}} - P_{\nu}^0\fn{\bs{P}_{\nu}}
				}^2.
\end{align}
We note that the third term on the right-hand side of the second equality vanishes by the definition of $P_{\nu}^0\fn{\bs{P}_{\nu}}$ in Eq.~\eqref{appeq: def P0nu} and the definition of $q_{\SD}\fn{\bs{p}_{\nu_I}}$ in Eq.~\eqref{appeq: def qSD}. We also note that $q_{\SD}\fn{\bs{P}_{\nu}}$ is typically small, and therefore it is treated as zero in the expression for the amplitude in the main text.

Next, for the contribution of the virtual propagation, we evaluate the momentum integral using the saddle-point approximation. The integral to be evaluated takes the following form:
\begin{align}
\int d^3  \bs{p}_{\nu_I}\,\exp\!\left[H_{0}\fn{\bs{p}_{\nu_I}}\right] R_{0} \fn{\bs{p}_{\nu_I}}.
\end{align}
where
\begin{align}
H_0 \fn{p^i_{\nu_I}}
	:=
	&
		- i\, P^0_{\nu}\fn{\bs{p}_{\nu_I}}\, \Delta T + i p_{\nu_I}^i L^i  \nonumber\\
	& 
		-\sumSandD {\sigma_{\SD} \over 2} \pn{\pn{-1}^{\SD} p_{\nu_I}^i - \ol{P^i}_{\SD0}}^2 \nonumber\\
	&
		- \sumSandD {\sigma_{t, \SD} \over 2}  
			\pn{\pn{-1}^{\SD} P^0_{\nu}\fn{\bs{p}_{\nu_I}}
				- \ol{P^0}_{\SD 0} - q_{\SD}\fn{\bs{p}_{\nu_I}}}^{2},
	\label{appeq: H0 defined}
\end{align}
and $R_{0} \fn{\bs{p}_{\nu_I}}$ denotes the non-exponential part associated with localization.

The first and second derivatives of Eq.~\eqref{appeq: H0 defined} can be evaluated in the same manner as in the case of the real propagation, as
\begin{align}
{\partial H_{0}\over\partial p_{\nu_I}^i}\fn{\bs{p}_{\nu_I}}
	&=- \sigma_{\tx{S}+ \tx{D}}\, \pn{p_{\nu_I}^i - P^i_{\nu} }
		-  i v^i_{\nu, \mr{vi}} \fn{\bs{p}_{\nu_I}}  \Delta T + i  L^i  + \cdots,
		\label{appeq: first derivative of H0}\\
{\partial^2 H_{0}\fn{\bs p_{\nu_I}}\over \partial p_{\nu_I}^i\partial p_{\nu_I}^j}
	&\simeq - \sigma_{\tx{S}+\tx{D}} \delta^{ij}.
			\label{appeq: second derivative of H0}
\end{align}

Therefore, the saddle point is given by
\begin{align}
p_{\nu_I \star}^i
	&:= P_{\nu}^i 
	+ i\,{L^i -  v^i_{\nu, \text{vi}} \Delta T 
			\over \sigma_{\tx{S}+ \tx{D}} }
	+ \mathcal{O}(\sigma^{-2}) .
\end{align}

The exponent \(H_0 \fn{p^i_{\nu_I}}\) evaluated at the saddle point is given by
\begin{align}
H_{0} \fn{\bs{p}_{\nu_I \star}}
	&=
		-  i P^0_{\nu}\fn{\bs{P}_{\nu_I}}  \Delta T + i P_{\nu}^i L^i \nonumber\\
	&\quad
		-{\sigma \over 2}  \pn{\ol{P^i}_{\tx{S}0}+ \ol{P^i}_{\tx{D}0}}^2 
		- {\sigma_{t} \over 2}  
				\pn{\ol{P^0}_{\tx{S}0} + \ol{P^0}_{\tx{D}0}
					- \pn{
						q_{\tx{S}}\pn{\bs{P}_{\nu}} +q_{\tx{D}}\pn{\bs{P}_{\nu}}
						} 
					}^2\nonumber \\
	&\quad
		-{1 \over 2\sigma_{\tx{S}+ \tx{D}}}  \pn{ L^i -  v^i_{\nu, \text{vi}}\Delta T}^2.
\end{align}

Based on the above results, after performing the integration, we find that the expressions for the amplitudes of the real propagation and the virtual propagation are given by Eqs.~\eqref{eq: amplitude on-shell} and \eqref{eq: amplitude off-shell}, respectively.

\section{Evaluation of the propagating time integration}
\label{app: Evaluation of the propagating time integration}
To derive the transition probability $P(\alpha \to \beta)$, one has to evaluate the propagation-time integral of the squared amplitude,
\begin{equation}
\label{appeq: probability given by time integration}
P(\alpha \to \beta)
	\propto \sum_{IJ} \int d \Delta T \,
	\mc A_{I,\alpha\beta}\,
	\mc A^{*}_{J,\alpha\beta};
\end{equation}
see footnote~\ref{footnote:Delta-T-integral} for the validity of this treatment.
As discussed in the main text, for practical purposes, it is sufficient to evaluate the propagation-time integral of the squared real-propagation contribution $\mc A^{\mr{re}}_{I,\alpha\beta}$ in the full amplitude,
\begin{align}
\label{appeq: flavor changing amplitude}
\mc A_{I,\alpha\beta} 
    = \mc A^{\mr{vi}}_{I,\alpha\beta} 
    + \theta\fn{\Delta T - \Delta T_{\mr{th},I}\fn{\bs{P}_{\nu}}} 
      \mc A^{\mr{re}}_{I,\alpha\beta}.
\end{align}
As mentioned in the main text, the contribution relevant to the oscillation formula is that of real propagation. Therefore, in the following, we focus on the contribution of $\mc A^{\mr{re}}_{I,\alpha\beta}$ to the amplitude. Since real propagation occurs over a classical length scale, we set the uncertainty factor associated with the intersection point of the vertices to unity, $\psi_X = 1$.

Then, what we need to evaluate is an integral of the following form:
\begin{align}
\int d\Delta T \,\exp\!\left[\DTexp\fn{\Delta T}\right].
\label{eq: Generic integral for propagating time}
\end{align}
The exponent is given by
\begin{align}
\label{I defined}
\DTexp\fn{\Delta T}
	&=   -{1 \over 2\sigma_{\tx{S}+ \tx{D}}}   \pn{ L^i -  v^i_{\nu_I} \fn{\bs{P}_{\nu}}\Delta T}^2
		-{1 \over 2\sigma_{\tx{S}+ \tx{D}}}   \pn{ L^i -  v^i_{\nu_J} \fn{\bs{P}_{\nu}}\Delta T}^2 \notag \\
	&\ \ \, 
    - i \,p_{\nu_I +}^0\fn{\bs{P}_{\nu}} \Delta T 
    + i \, \pn{p_{\nu_J +}^0\fn{\bs{P}_{\nu}}}^* \Delta T.
\end{align}

The first and second derivatives of the exponent are given by 
\begin{align}
{\partial{\DTexp}\fn{\Delta T}\over\partial \Delta T}
	&= 
		- i \pn{ p_{\nu_I +}^0\fn{\bs{P}_{\nu}} -  (p_{\nu_J +}^0\fn{\bs{P}_{\nu}})^*
        } \nonumber \\
	&\quad
		+ \sigma_{\tx{S}+ \tx{D}}^{-1} v^i_{\nu_I}  \pn{ L^i -  v^i_{\nu_I} \fn{\bs{P}_{\nu}}\Delta T} 
		 + \sigma_{\tx{S}+ \tx{D}}^{-1} v^i_{\nu_J}  \pn{ L^i -  v^i_{\nu_J} \fn{\bs{P}_{\nu}}\Delta T} \nonumber\\
	&\simeq  - \sigma_{\tx{S}+ \tx{D}}^{-1} \pn{ \sumIandJ \pn{v^i_{\nu_{X}}}^2}
		\pn{
		\Delta T - { \sumIandJ L^i v^i_{\nu_{X}} \over \sumIandJprime \pn{v^i_{\nu_{X'}}}^2} 
		+ i { 
        E_{\nu_I}\fn{\bs{P}_{\nu}} -  E_{\nu_J}\fn{\bs{P}_{\nu}}
			 \over \sigma_{\tx{S}+ \tx{D}}^{-1}  \sumIandJ \pn{v^i_{\nu_{X}}}^2}
		} \notag \\
    &=  - \sigma_{\tx{S}+ \tx{D}}^{-1} \pn{ \sumIandJ \pn{v^i_{\nu_{X}}}^2}
		\pn{
		\Delta T - \Delta T_{0,IJ}
		+ i \sigma_{\tx{S}+ \tx{D}} { 
        E_{\nu_I}\fn{\bs{P}_{\nu}} -  E_{\nu_J}\fn{\bs{P}_{\nu}}
			 \over  \sumIandJ  \pn{v^i_{\nu_{X}}}^2}
		},
		\label{appeq: first derivative of I}\\
{\partial^2 \DTexp\fn{\Delta T}\over \partial \pn{\Delta T}^2 }
	&=	-
		\sigma_{\tx{S}+ \tx{D}}^{-1}  \sumIandJ \pn{v^i_{\nu_{X}}}^2,
			\label{appeq: second derivative of I}
\end{align}
where we have retained only the terms that survive at leading order in $\Gamma_{\nu_I}$, and we have introduced
\begin{align}
\ol{v}_{IJ}^i 
	&:= {1 \over 2} \pn{v^i_{\nu_{I}} + v^i_{\nu_{J}}}, \\
\Delta v_{IJ}^i 
	&:= {1 \over 2} \pn{v^i_{\nu_{I}} - v^i_{\nu_{J}}}, \\
\Delta T_{0,IJ}
		& := 
            { \sumIandJ L^i v^i_{\nu_{X}} \over \sumIandJprime \pn{v^i_{\nu_{X'}}}^2}
            =
            {2 L^i \bar{v}_{IJ}^i
			\over \pn{v_{\nu_I}^i}^2  + \pn{v_{\nu_J}^i}^2}.
\end{align}

Here, since the neutrino mass $m_{\nu_I}$ is sufficiently small, we can expand
\begin{align}
v^i_{\nu_{I}}
	&\simeq \ol{v}_{0}^i \pn{1 - {m_{\nu_I}^2 \over 2 |\bs{P}_{\nu}|^2}},&
\ol{v}_{0}^i 
    &:= {P^i_{\nu} \over |\bs{P}_{\nu}|}.&
    \label{eq:vnuI-expansion}
\end{align}
It is necessary to draw attention once again to the approximate treatment employed in this manuscript.
As mentioned in Footnote No.~\ref{footnote:Delta-T-integral}, in our formalism, as a matter of principle, the most probable values of all intermediate states are determined by the configurations of the external particles.
The most probable form of the spatial momentum of the intermediate states $\bs{P}_{\nu_I}$, it is given as $\bs{P}_{\nu_I} \simeq \bs{P}_{\nu}$, where the form of $\bs{P}_{\nu}$ is given in Eq.~\eqref{eq:P-spatial_nu}.

It is clear that $\ol{v}_{IJ}^i$ and $\Delta T_{0,IJ}$ are zeroth order in the squared-mass difference, $\Delta m_{IJ}^2$, whereas $\Delta v_{IJ}^i$ is first order in the squared-mass difference. 
For the later convenience, we show their expansions:
\begin{align}
\ol{v}_{IJ}^i 
	&\simeq \ol{v}_{0}^i \pn{1 - {(m_{\nu_I}^2 + m_{\nu_J}^2)  \over 4 
        |\bs{P}_{\nu}|^2}} \simeq \ol{v}_{0}^i, \\
\Delta v_{IJ}^i 
	&\simeq - \ol{v}_{0}^i 
    {
    \Delta m_{IJ}^2   \over 4 |\bs{P}_{\nu}|^2
    }, \\
\Delta T_{0,IJ}
		& \simeq \Delta T_{0}, \\
\Delta T_{0} 
        & := {L^i \bar{v}_{0}^i \over \pn{\bar{v}_{0}^i}^2}.
\end{align}
Here, the value of $\Delta T_{0} $ defined above plays a crucial role in the discussion regarding whether or not on-shell channels open.
We note that
\begin{align}
\sumIandJ \pn{v^i_{\nu_{X}}}^2
    &\simeq 2 \pn{\ol{v}_{0}^i}^2.
\end{align}

The saddle point of propagating time $\Delta T_{\star,IJ}$ is then given by
\begin{align}
\Delta T_{\star, IJ}
	\simeq  \Delta T_{0,IJ}
		- i \sigma_{\tx{S}+ \tx{D}}
            { E_{\nu_I}\fn{\bs{P}_{\nu}} -  E_{\nu_J}\fn{\bs{P}_{\nu}}
			 \over 2 \pn{\ol{v}_{0}^i}^2 }.
\end{align}
We note that the real part of $\Delta T_{\star, IJ}$ is zeroth order in the squared-mass difference, $\Delta m_{IJ}^2 $, whereas its imaginary part is first order.
Therefore, evaluating the exponent at the saddle point, and expanding it up to second order in $\Delta m_{IJ}^2$, we obtain
\begin{align}
\label{appeq: ITstar evaluation}
\DTexp\fn{\Delta T_{\star, IJ}} 
	&=	\DTexp\fn{\Delta T_{0,IJ}}
		+{\partial{\DTexp}\fn{\Delta T_{0,IJ}}\over \partial \Delta T}  
		i\im[\Delta T_{\star, IJ}]
	    +{1\over2} {\partial^2 \DTexp\fn{\Delta T}\over \partial \pn{\Delta T}^2 } 
		\pn{i\im[\Delta T_{\star, IJ}]}^2\\
	&\simeq
		- \sigma^{-1}_{\tx{S}+\tx{D}} \pn{L^i - \ol{v}_0^i \Delta T_0}^2
		-  \sigma^{-1}_{\tx{S}+\tx{D}}\pn{\Delta v_{IJ}^i}^2 \Delta T_0^2
		- i \pn{p_{\nu_I +}^0\fn{\bs{P}_{\nu}} -  (p_{\nu_J +}^0\fn{\bs{P}_{\nu}})^* }\Delta T_0\\
	&\qquad
		-{1 \over 2} {\sigma_{\tx{S}+ \tx{D}} \over 2  \pn{\bar{v}_0^i}^2} \pn{ E_{\nu_I}\fn{\bs{P}_{\nu}} -  E_{\nu_J}\fn{\bs{P}_{\nu}}}^2  \\
	&\simeq
		-  \sigma^{-1}_{\tx{S}+\tx{D}}\,
		\pn{L^i  \hat{P}_{\ol{v}_{0}}^{ij} 
			}^2
		-  \sigma^{-1}_{\tx{S}+\tx{D}}\,
				\pn{
				L^i \ol{v}^i_{0} \over \ol{v}^2_{0}
				}^2
			{\pn{\Delta m_{IJ}^2 }^2 \over 16  |\bs{P}_{\nu}|^4}
		- \pn{
			 i{\Delta m_{IJ}^2  \over 2 |\bs{P}_{\nu}| } 
             + {m_{\nu_I}\Gamma_{\nu_I} + m_{\nu_J}\Gamma_{\nu_J} \over 2 |\bs{P}_{\nu}| }
			 }
			 {
				L^i \ol{v}^i_{0} \over \ol{v}^2_{0}
				},
\end{align}
where
\begin{align}
\hat{P}_{\ol{v}_{0}}^{ij} 
	&:= \delta^{ij} -
				{
				 	\ol{v}_{0}^i  \ol{v}_{0}^{j} \over \ol{v}_{0}^2
					}.
\end{align}
For both the terms linear and quadratic in $L^i$, we have retained only the leading contributions in $\Delta m_{IJ}^2$ or $\Gamma_{\nu_I}$. As a result, the coefficient of the term linear in $L^i$ is first order in $\Delta m_{IJ}^2$ or $\Gamma_{\nu_I}$, while the coefficient of the term quadratic in $L^i$ is second order in $\Delta m_{IJ}^2$.

The first term in Eq.~\eqref{appeq: ITstar evaluation} shows that the exponential factor selects the propagation direction,
\begin{align}
L^i_{\mr{prop}} := \frac{\ol{v}_0^i}{\sqrt{\left(\ol{v}_0^i\right)^2}} \, L ,
\end{align}
by exponentially suppressing contributions proportional to all other directions of $L^i$.
We remind readers of the expansion in Eq.~\eqref{eq:vnuI-expansion}, where for active neutrinos, $v^i_{\nu_I}$ is very close to $\ol{v}_{0}^i$.
The second term suppresses the oscillation probability when
\begin{align}
L \gtrsim 
L^{\mathrm{coh}}_{IJ}
    &:= 4{\sqrt{\sigma_{\tx{S}+\tx{D}}} \over \Delta m_{IJ}^2}
        \left|\bs{P}_{\nu}\right|^2
\end{align}
is satisfied.
The third term describes flavor oscillations and the decoherence effect due to intermediate particle decay. From this term, the oscillation length can be read off as
\begin{align}
L^{\mathrm{osc}}_{IJ}
    &:= {4 \pi \left|\bs{P}_{\nu}\right| \over \Delta m_{IJ}^2},
\end{align}
and the oscillation probability is suppressed when
\begin{align}
L \gtrsim
    L^{\mr{dec}}_{IJ}
    &:= {2\left|\bs{P}_{\nu}\right| \over m_{\nu_I}\Gamma_{\nu_I} + m_{\nu_J}\Gamma_{\nu_J}},
\end{align}
is satisfied.

Here, as previously noted, although $\bs{P}_\nu$ is fundamentally a degree of freedom distinct from $\bs{L}$, if we posit, on phenomenological grounds, that $\bs{P}_\nu$ is parallel to $\bs{L}$ (and, consequently, that $\ol{\bs{v}}_0$ is also parallel to $\bs{L}$), it can be immediately verified that the leading term reproduces the standard form of neutrino oscillation shown in Eq.~\eqref{eq: Standard Formula}.
In this paper, we will not discuss corrections arising from the fact that $\bs{P}_\nu$ and $\bs{L}$ are not perfectly parallel.
This matter will be addressed in our subsequent future work.

\end{widetext}


\bibliographystyle{utphysmod}
\bibliography{ref}

\end{document}